\documentclass[acmsmall]{acmart}

\usepackage{graphicx}
\usepackage{amsmath,amssymb}
\usepackage{mathrsfs}
\usepackage{bm}
\usepackage{booktabs}
\usepackage{multirow}
\usepackage{makecell}
\usepackage{subfigure}
\usepackage{algorithm}  
\usepackage{algpseudocode}
\usepackage[figuresleft]{rotating}
\AtBeginDocument{%
  \providecommand\BibTeX{{%
    \normalfont B\kern-0.5em{\scshape i\kern-0.25em b}\kern-0.8em\TeX}}}

\setcopyright{acmcopyright}
\copyrightyear{2018}
\acmYear{2018}
\acmDOI{10.1145/1122445.1122456}

\acmJournal{JACM}
\acmVolume{37}
\acmNumber{4}
\acmArticle{111}
\acmMonth{8}

\begin{document}

\title{Deep Meta-learning in Recommendation Systems: A  Survey}

\author{Chunyang Wang, Yanmin Zhu, Haobing Liu, Tianzi Zang, Jiadi Yu, Feilong Tang }
\email{wangchy@sjtu.edu.cn, yzhu@sjtu.edu.cn, liuhaobing@sjtu.edu.cn, zangtianzi@sjtu.edu.cn, jiadiyu@sjtu.edu.cn, tang-fl@cs.sjtu.edu.cn}
\affiliation{%
	\institution{Shanghai Jiao Tong University}
	\city{Shanghai}
	\country{China}
}
%
%
%
%

\begin{abstract}
  Deep neural network based recommendation systems have achieved great success as information filtering techniques in recent years. However, since model training from scratch requires sufficient data, deep learning-based recommendation methods still face the bottlenecks of insufficient data and computational inefficiency. Meta-learning, as an emerging paradigm that learns to improve the learning efficiency and generalization ability of algorithms, has shown its strength in tackling the data sparsity issue. Recently, a growing number of studies on deep meta-learning based recommenddation systems have emerged for improving the performance under recommendation scenarios where available data is limited, e.g. user cold-start and item cold-start. Therefore, this survey provides a timely and comprehensive overview of current deep meta-learning based recommendation methods. Specifically, we propose a taxonomy to discuss existing methods according to recommendation scenarios, meta-learning techniques, and meta-knowledge representations, which could provide the design space for meta-learning based recommendation methods. For each recommendation scenario, we further discuss technical details about how existing methods apply meta-learning to improve the generalization ability of recommendation models. Finally, we also point out several limitations in current research and highlight some promising directions for future research in this area.
 
\end{abstract}



\maketitle
\section{Introduction}
In recent years, recommendation systems working as filtering systems for alleviating information overload have been widely applied in various online applications including e-commence, entertainment services, news, and so on.  By presenting personalized suggestions among a large number of candidates, recommendation systems have achieved great success in improving user experience and increasing the attractiveness of online platforms. With the development of data-driven machine learning algorithms \cite{CF_survey, CF_1}, especially deep learning based methods \cite{zhang2019deep, NCF, WideDeep}, academic and industrial research in this field has greatly improved the performance of recommendation systems in terms of accuracy, diversity, interpretability, and so on.

Due to expressive representation learning abilities to discover hidden dependencies from sufficient data, deep learning based methods have been largely introduced in contemporary recommendation models \cite{zhang2019deep, Graph_survey}. By leveraging a great number of training instances with diverse data structures (e.g., interaction pairs \cite{zhang2019deep}, sequences\cite{Sequential_Survey}, and graphs \cite{Graph_survey}), recommendation models with deep neural architectures are usually designed to effectively capture nonlinear and nontrivial
user/item relationships. However, conventional deep learning based recommendation models are usually trained from scratch with sufficient data based on predefined learning algorithms. For instance, the regular supervised learning paradigm typically
trains a unified recommendation model with interactions collected from all users and performs recommendation over unseen interactions based on learned feature representations. Such deep learning based methods are usually data-hungry and computation-hungry. In other words, the performance of deep learning based recommendation systems heavily relies on the availability of a great amount of training data and sufficient computation. In practical recommendation applications, data collection mainly originates from users' interactions observed during their visits to online platforms. There exist recommendation scenarios where available user interaction data is sparse (e.g. cold-start recommendation) and computation for model training is restrained (e.g. online recommendation). Consequently, both data insufficiency and computation inefficiency issues bottleneck deep learning based recommendation models.

Recently, meta-learning provides an appealing learning paradigm that focuses on strengthening the generalization ability of machine learning methods against the insufficiency of data and computation \cite{ML_survey_1, ML_survey_2}. The key idea of meta-learning is to gain prior knowledge (named \emph{meta-knowledge}) about efficient task learning from previous learning processes of multiple tasks. Then, the meta-knowledge could help facilitate fast learning over new tasks, which is supposed to have good generalization performance on unseen tasks. Here, a task usually refers to a set of instances belonging to the same class or having the same property, involving an individual learning process on it. Different from improving the representation learning capacity of deep learning models, meta-learning focuses on learning better learning strategies to substitute for fixed learning algorithms, known as the concept of \emph{learn to learn}. Due to its great potential for fast adaptation over unseen tasks, meta-learning techniques have been applied in a wide range of research domains including image recognition \cite{cai2018memory, zhu2020multi}, image segmentation \cite{image_survey}, natural language processing \cite{NLP_meta_survey_1}, reinforcement learning \cite{Rl_meta_1, Rl_meta_2} and so on.

The benefits of meta-learning are well-aligned with the need of promoting recommendation models over scenarios suffering from limited instances and inefficient computation. Early efforts on meta-learning based recommendation methods mainly fall into personalized recommendation algorithm selection \cite{ren2019survey, Survey_selection}, which extracts meta dataset features and selects suitable recommendation algorithms for different datasets (or tasks). Though applying the idea of extracting meta-knowledge and generating task-specific models, this definition of meta-learning is closer to studies in automated machine learning \cite{AutoML_survey_1, AutoML_survey_2}. Afterward, deep meta-learning \cite{DML_survey} or neural network meta-learning \cite{ML_survey_2} emerges and gradually become the mainstream of meta-learning techniques typically discussed in the recommendation models \cite{MeLU, Meta-Embedding}. As introduced in \cite{DML_survey, ML_survey_2}, \emph{Deep Meta-Learning} aims to extract meta-knowledge to allow for fast learning of deep neural networks, which brings enhancement to the currently popular deep learning paradigm. Since 2017, deep meta-learning  has gained attention in the research community of recommendation systems. Advanced meta-learning techniques are firstly applied to alleviate data insufficiency (i.e., cold-start issue) when training conventional deep recommendation models. For example, the most successful optimization-based meta-learning framework MAML which learns meta-knowledge in the form of parameter initialization of neural networks firstly shows great effectiveness in the cold-start recommendation scenario \cite{MeLU}. Besides that, diverse recommendation scenarios such as click-through-rate prediction \cite{Meta-Embedding}, online recommendation \cite{SML}, and sequential recommendation \cite{Mecos} are also studied under the meta-learning schema, to improve the learning ability in the setting of data insufficiency and computation inefficiency.

In this paper, we provide a timely and comprehensive survey of the rapidly growing studies of deep meta-learning based recommendation systems. As we investigated, although there have been some surveys on meta-learning or deep meta-learning that summarize details of general meta-learning methods and their applications \cite{ML_survey_1, DML_survey, ML_survey_2}, there still lacks attention to recent advances in recommendation systems. In addition, there are several surveys on meta-learning methods in other application domains, such as Natural Language Processing \cite{NLP_meta_survey_1, NLP_meta_survey_2}, Multimodality \cite{Multimodality_Meta_survey} and Image Segmentation \cite{image_survey}. However, no previous survey centers on the deep meta-learning in recommendation systems. Compared with them, our survey is the first attempt to fill the gap, providing a systematic review of up-to-date papers on the combination of meta-learning and recommendation systems.

In our survey, we aim to thoroughly review the literature on the deep meta-learning based recommendation systems, which can benefit readers and researchers for a comprehensive understanding of this topic. To carefully position works in this field, we provide a taxonomy with three perspectives including recommendation scenarios, meta-learning techniques, and meta-knowledge representations. Moreover, we mainly discuss related methods according to recommendation scenarios and present how different works utilize meta-learning techniques to extract specific meta-knowledge with diverse forms such as parameter initialization, parameter modulation, hyperparameter optimization, .etc. We hope our taxonomy could provide a design space for developing new deep meta-learning based recommendation methods. In addition, we also summarize common ways for meta-learning task construction which is a necessary setup of the meta-learning paradigm. 

The structure of this survey is organized as follows. In Section~\ref{sec:2}, we introduce the common foundations of meta-learning techniques and typical recommendation scenarios in which meta-learning methods have been studied to alleviate data insufficiency and computation inefficiency.  In Section~\ref{sec:3}, we present our taxonomy consisting of three independent axes. In Section~\ref{sec:4}, we summarize different ways of meta-learning recommendation task construction used in the literature. Then we elaborate on methodological details of existing methods applying meta-learning techniques in different recommendation scenarios in Section~\ref{sec:5}. Finally, we discuss promising directions for future research in this field in Section~\ref{sec:6} and conclude this survey in Section~\ref{sec:7}.

\textbf{Paper Collection.}  We summarize over 50 high-quality papers which are highly related to deep meta-learning based recommendation systems. We carefully retrieve these papers using Google Scholar and DBLP as main search engines with major keywords including meta-learning + recommendation, meta + recommendation, meta + CTR, meta + recommender, etc. We particularly pay attention to top-tier conferences and journals including KDD, SIGIR, WWW, AAAI, IJCAI, WSDM, CIKM, ICDM, TKDE, TKDD, TOIS, so as to ensure that high-profile papers are covered.

\section{Foundations}
\label{sec:2}
In this section, we present the necessary foundations for discussing deep meta-learning based recommendation methods. Firstly, we summarize the core ideas and representative works of different categories of meta-learning techniques. Afterward, we introduce typical recommendation scenarios in which meta-learning techniques have been studied and applied.

\subsection{Meta-learning} 
To comprehensively understand the concept of meta-learning, we first formalize the paradigm of meta-learning and contrast the conventional machine learning paradigm with the meta-learning paradigm in detail. Then, we briefly present three mainstreams of meta-learning techniques, including \emph{metric-based}, \emph{model-based} and \emph{optimization-based} meta-learning techniques by summarizing their core ideas and introducing several typical related works. For convenience, we list some general symbols and their descriptions in Table \ref{tab1}.

\begin{table}[t]
	\caption{Notations used in this paper.}\label{tab1}
	\centering
	\begin{tabular}{cl}
		\toprule
		\textbf{Notations} & \textbf{Descriptions} \\
		\midrule
		$u_i $ & User $i$ \\
		$v_j $ & Item $j$ \\
		$r_{u_i, v_j}$  & An interaction between $u_i$ and $v_j$ (explicit rating or implicit feedback) \\
		$\bm{x}_k, y_k $ &  Representation and label of $k$-th instance (e.g. an interaction)\\
		$\mathcal{T}_{i}$ & $i$-th recommendation task \\
		$\mathcal{S}_{i}$ & Support set of a task $\mathcal{T}_{i}$ \\
		$\mathcal{Q}_{i}$ & Query set of a task $\mathcal{T}_{i}$ \\
		$\mathcal{D}^{train}$ & Meta-training dataset \\
		$\mathcal{D}^{test}$ & Meta-testing dataset \\
		$f_\theta$ & Base recommendation model/function \\
		$\theta$ & Parameters of the base recommendation model \\ 
		$\theta_{\mathcal{T}_{i}}$ & Task-specific parameters of a personalized model for $\mathcal{T}_{i}$\\ 
		$\alpha $  & Local update rate in the optimization-based meta-learning \\
		$\beta $  & Global update rate in the optimization-based meta-learning \\
		$\mathcal{L}(f_\theta, *)$ & Loss function of the base recommendation model over a given dataset\\
		$\mathcal{F}_\omega$ & Meta-learner parameterized with $\omega$ \\
		$\omega$ & Meta-knowledge obtained with the meta-learner \\
		\bottomrule
	\end{tabular} 
\end{table}

\subsubsection{Formalizing Meta-learning} As commonly understood as the concept of \emph{learning to learn}, meta-learning mainly contributes to improving the generalization ability of base learning models or algorithms, so as to learn new tasks better or more quickly. Generally, the core idea of meta-learning paradigm is learning prior knowledge, i.e. \emph{meta-knowledge}, across multiple tasks, where each task refers to a learning process which tries to perform well on its own instances. The learning processes of different tasks are treated as training instances observed by meta-learning methods. By defining the form of meta-knowledge and extracting meta-knowledge across multiple existing tasks, meta-learning methods enable the learning processes of new tasks to be more effective.

Formally, in the training phase of meta-learning paradigm, we assume that a set of training tasks sampled from a task distirbution $p(\mathcal{T})$ are available as a meta-training dataset which is denoted as $\mathcal{D}^{train} = \{\mathcal{T}_{i}\}_{i=1}^{M}$. All instances $\mathcal{D}_i$ of a task $\mathcal{T}_{i}$ consists of its own training instances denoted as the support set $\mathcal{S}_{i}$ and evaluation instances denoted the query set $\mathcal{Q}_{i}$. Take a task under the supervied learning scheme as example. Given the support set $\mathcal{S}_{i} =\{(\bm{x}_k,y_k)\}_{k=1}^{S}$ consisting of $S$ training instances, the task $\mathcal{T}_{i}$ aims to learn a mapping function(or model) $f_{\theta}: \mathcal{X} \rightarrow \mathcal{Y}$ by minimizing the empirical loss $\mathcal{L}(f_{\theta}, \mathcal{S}_{i})$. The task-specific parameters $\theta_{\mathcal{T}_{i}}$ of the mapping function for the task $\mathcal{T}_{i}$ is obtained as follows: 
\begin{equation} 
\theta_{\mathcal{T}_{i}} = \mathop{\arg\min}\limits_{\theta} \mathcal{L}(f_{{\theta}}, \mathcal{S}_{i})
\end{equation}
where the loss function $\mathcal{L}(f_{\theta},*)$ could be a cross-entropy loss for classification tasks or regression loss such as mean squared error for regression tasks. Note that the training process of each task is usually conducted the same as the regular supervised learning. To measure the generalization performance of the trained model over unseen instances, a set of evaluation instances $\mathcal{Q}_{i} =\{(\bm{x}_k,y_k)\}_{k=1}^{Q}$ are sampled from the same distribution of the task $\mathcal{T}_{i}$. The learned mapping function $f_{\theta_{\mathcal{T}_{i}}}$ is supposed to perform well by investigating the empirical loss $\mathcal{L}(f_{\theta_{\mathcal{T}_{i}}}, \mathcal{Q}_{i})$ or other evaluation metrics in different settings. To be mentioned, learning tasks in other schemes such as reinforcement learning \cite{reinforcement_meta} and unsupervised learning \cite{unsupervised_ML} are also studied.

In the training processes of different meta-training tasks in $\mathcal{D}^{train}$, even if the form of the mapping functions could be the same, how to learn task-specific models is still distinct and guided by learnable settings about task learning. For example, approximating using neural networks with the same structure requires suitable hyper-parameters or initialization settings which are likely to be different for different tasks. In other words, the learning of each task $\mathcal{T}_{i}$ also depends on \emph{how to learn}, which is defined as meta-knowledge $\omega$ under the meta-learning paradigm. Therefore, the task-specific learning of $\mathcal{T}_{i}$ could be formalized as follows: 
\begin{equation} 
\theta_{\mathcal{T}_{i}} = h_\omega(f_{\theta},\mathcal{T}_{i},\mathcal{L})
\end{equation}
where $h_\omega(*)$ denotes the meta-learning approches of utilizing meta-knowledge to ensure effective learning of task $\mathcal{T}_{i}$ with the same mapping function $f_{\theta}$ and loss function $\mathcal{L}$. 

Instead of assuming the meta-knowledge $\omega$ is pre-defined and fixed for all tasks, meta-learning allows for learning $\omega$ to enable each task to be learned better. Manually searching in the whole meta-knowledge space is impractical in most cases. The goal of meta-learning is to learn the optimal $\omega$ which could be utilized to guide task-specific learning of all tasks to perform better. Formally, given all training tasks $\mathcal{D}^{train} = \{\mathcal{T}_{i}\}_{i=1}^{M}$, the optimal meta-knowledge $\omega^*$ are obtained as follows:  
\begin{equation} 
\omega^* = \mathop{\arg\min}\limits_{\bm{\omega}} \sum_{\mathcal{T}_{i} \in \mathcal{D}^{train}} \mathcal{L}(f_{\theta_{\mathcal{T}_{i}}}, \mathcal{Q}_{i}) = \mathop{\arg\min}\limits_{\bm{\omega}} \sum_{\mathcal{T}_{i} \in \mathcal{D}^{train}} \mathcal{L}(f_{h_\omega(f_{\theta},\mathcal{T}_{i},\mathcal{L})}, \mathcal{Q}_{i})
\end{equation}
where the objective of training meta-learning methods is to observe better performance (e.g., lower empirical loss) over the corresponding query set $\mathcal{Q}_{i}$ of each task. Note that, the meta-knowledge is learned across multiple tasks since it is supposed to mine across-task characteristics of different task learning processes and has great generalization ability against the task differences.

In contrast with conventional machine learning, (e.g., regular supervised learning paradigm), the meta-learning paradigm mainly has the following properties: 1) \textbf{Learning objective}. The learning objective of meta-learning, i.e., \emph{meta-optimization objecive}, is to facilitate the learning over unseen tasks, while conventional machine learning aims to facilitate the learning over unseen instances of the same task. 2) \textbf{Setup of task division}. For regular supervised machine learning, all instances are usually sampled from the data distribution of a single task. There are also multi-task learning \cite{multi_task_survey} or transfer learning frameworks \cite{transfer_learning_survey} which consider knowledge transfer across multiple tasks. However, these frameworks mainly consider a pair of tasks or a small number of known tasks, and transfer knowledge from other tasks as additional information, such as pretraining techniques or joint optimization strategies. In comparison, under the meta-learning paradigm, a larger number of tasks with relatively fewer instances are explicitly split according to specific properties (e.g., classes, attributes, or time), so as to extract prior knowledge about task learning at a higher level, i.e., learn to learn. 3) \textbf{Learning framework}. A common framework of meta-learning follows a bi-level learning structure consistent with meta-optimization objectives. The inner-level learning focuses on task-specific learning to generate training instances of the outer-level learning. The outer-level learning is responsible for learning the meta-knowledge across multiple instances. For most regular machine learning, only one level of learning is conducted over all supervised instances through batch learning, which is the same as the inner-level learning in the meta-learning paradigm.

\subsubsection{Mainstream Frameworks of Meta-learning Techniques}
As summarized by previous meta-learning surveys \cite{ML_survey_1, DML_survey}, meta-learning techniques mainly fall into three categories, named metric-based, model-based, and optimization-based meta-learning methods. Next, we will elaborate on the formalization, technical details, and representative works of each category and discuss their pros and cons compared with each other.

\textbf{Metric-based Meta-learning} resorts to the idea of metric learning and mainly represents meta-knowledge $\omega$ in the form of a meta-learned feature space where the similarity of support instances and query instances are compared. Specifically, task-specific learning in metric-based techniques is conducted in the form of non-parametric learning. In other words, in the inner-level learning of each task, the parameters of the mapping function $f_{\theta}$ are not optimized to fit the training instances $\mathcal{S}_{i}$ but directly utilized to generate labels of evaluation instances $\mathcal{Q}_{i}$.  For the  mapping function $f_{\theta}$, metric-based methods mainly rely on a similarity scoring function $sim(\bm{x}_i,\bm{x}_j)$ which takes embeddings of two instances (e.g., a training (support) instance and a evaluation (query) instance) as inputs and calculates a similarity weight in the meta-learned feature space. Then the label of an evaluation instance is assigned by the weighted combination of labels from all training (support) instances. Formally, the predicted label vector $\hat{\bm{y}}_i$ of a query instance $\bm{x}_i$ in the task $\mathcal{T}_{i}$ could be obtained as follows: 
\begin{equation} 
\hat{\bm{y}}_i = \sum\limits_{(\bm{x}_j,\bm{y}_j) \in \mathcal{S}_{i}} sim(\bm{x}_i,\bm{x}_j) \bm{y}_j
\end{equation}
Note that we simply present a basic form of metric-based meta-learning. In the litertature, the similarity function $sim(\bm{x}_i,\bm{x}_j)$ and label generating could be achieved in different forms such as siamese nets \cite{Siamese_nets}, matching nets \cite{Matching_nets}, prototypical nets \cite{Prototypical_nets}, relation nets \cite{Relation_nets}, and graph neural networks \cite{GraphNN_nets}. 

For outer-level learning, metric-based meta-learning aims to learn the feature space for effectively comparing instance similarity in new tasks. Therefore, the meta-knowledge $\omega$ coincides with the parameters $\theta$ in the mapping function of the inner-level learning. Then $\theta$ are optimized by minimizing the empirical loss over query set of multiple training tasks as equation 3. To be mentioned, the $\theta_{\mathcal{T}_{i}}$ is the same as the $\bm{\theta}$ since the inner-level task-specific learning is non-parametric.

\textbf{Model-based Meta-learning} is another widely used meta-learning technique with the help of the powerful representation ability of neural network structures. The key idea of model-based methods is to meta-learn a model or a module to encode the internal states of a task by observing its support instances. Conditioned on the internal states, the model-based meta-learner could capture task-specific information and guide task-adaptive predictions for evaluation instances.

In the model-based meta-learning, inner-level learning mainly focuses on encoding the support instances (or gradients) of the task into representations of the internal state with a neural network structured model such as feed-forward networks, recurrent neural networks \cite{Recurrent_Model, Recurrent_Model_2}, convolutional neural networks \cite{SNAIL} or hypernetworks \cite{hypernet_1, hypernet_2}. The predictions of query instances are usually obtained with a modulated predictor conditioned on the encoded task-specific state representation.  Formally, the prediction of a query instance $\bm{x}_i$ in the task $\mathcal{T}_{i}$ could be obtained as follows: 
\begin{equation} 
\hat{\bm{y}}_i = f_{g_\omega(\theta,\mathcal{D}_i)}(\bm{x}_i)
\end{equation}
where the meta-knowledge $\omega$ plays a role in mapping task-specific states to modulation signals to predictors or optimization strategies. In general, $\omega$ is represented in the form of a external meta model $g$. The meta model $g$ could be instantiated with nerual networks \cite{wang2016learning} or external memories \cite{santoro2016meta}. For the outer-level learning, the optimization of the meta-learner is usually coupled into the training of the inner-level mapping function since the outputs of the inner-level learning relies on the outputs of the meta-learner.

\textbf{Optimization-based Meta-learning} strictly follows a bi-level optimization structure and separates the inner-level learning and outer-level learning via different gradient descent steps. We take a famous framework namly model-agnostic meta-learning (MAML) as an example. These are many studies extend the MAML frameworks. Specifically, in the inner-level learning, a base model performs as the predictor and conducts a few steps of local optimization based on the emperical loss over support instances as follows: 
\begin{equation} 
\theta_{\mathcal{T}_{i}}  = \theta - \gamma \nabla_\theta \mathcal{L}(f_\theta, \mathcal{S}_{i} ).
\end{equation}
where $\theta$ is the initialization of the base model parameters. We simply show one step of gradient descent. By performing the local update of the base model, $\theta_{\mathcal{T}_{i}} $ is utilized as the learned model after task-specific learning of the task $\mathcal{T}_{i}$. Here, task-specific learning refers to regular gradient descent based optimization, which is also the reason why this category is called optimization-based meta-learning. 

The meta-knowledge $\omega$ is represented in the form of parameter initialization in MAML, i.e., $\theta$. There are also other types of representation of meta-knowledge been studied. The $\theta$ is assigned to each task as meta-learned global initialization before task-specific learning. Therefore, in the outer-level learning, the $\theta$ is optimized by minimizing evaluation loss across different tasks to ensure that the initialization has generalization capacity as the meta-knowledge. Formally, the outer-level optimization, i.e., meta-optimization is conducted as follows: 

\begin{equation} 
\theta \leftarrow \theta - \alpha \nabla_{\theta} \sum\nolimits_{\mathcal{T}_{i} \in \mathcal{D}^{train}} \mathcal{L}(f_{\theta_{\mathcal{T}_{i}} }, \mathcal{Q}_{i} )
\end{equation} 
where the global intialization $\theta$ is updated across all tasks in the meta-training dataset $\mathcal{D}^{train}$ with second-order gradients, since $\theta_{\mathcal{T}_{i}} $ is obainted with gradient descents as equation (6).

\textbf{Discussion: Pros and Cons} These three frameworks of meta-learning techniques discussed above roughly covers most of the existing meta-learning methods. We conclude their advantages and disadvantages in terms of computation efficiency, the sensitivity of task distribution, and applicability. First, metric-based meta-learning has a small computational burden since simple similarity calculation requires no additional task-specific model update over new tasks. However, when task distribution is complex, metric-based methods usually perform unstably in the meta-test phase since no task information is absorbed to cope with task differences. Second, model-based meta-learning has relatively simple optimization steps compared with optimization-based meta-learning which requires second-order gradients. In addition, developed with diverse neural network structures, model-based methods usually have broader applicability compared to the other two. However, this category is criticized to perform worse over out-of-distribution tasks, i.e., is sensitive to task distribution. Third, the key advantage of optimization-based meta-learning is that it is usually agnostic to base model structure and could be compatible with diverse based models. In practice, optimization-based meta-learning show better generalization ability when task distribution is complex. However, this category of methods mainly suffers from heavy computation due to two levels of gradient descents.  

\subsection{Recommendation Scenarios } 

In the following, we will introduce typical scenarios of recommender systems that have been studied from the perspective of meta-learning, including cold-start recommendation,  click-through rate prediction, online recommendation, point-of-interest recommendation, and sequential recommendation. There also exist sporadic studies discussing meta-learning in other recommendation scenarios (e.g. cross-domain recommendation, multi-behavior recommendation), but we only give them a brief introduction when dicussing concrete methods in Section~\ref{sec:5}.

\textbf{Cold-start recommendation.} Despite the successful development of deep learning in general recommendation methods, one critical challenge to be addressed is the cold-start problem. Typically, the data scarcity issue commonly exists under the cold-start situations where new users come to visit the online platforms or new items are presented. Because observed user-item interactions are usually limited, traditional collaborative filtering methods \cite{CF_1} or deep learning methods \cite{NCF, WideDeep}, which require abundant training data are hard to perform well. Instead of being confined to interaction records, content-based methods depict users and items based on diverse content information, such as user and item attributes \cite{cold_start_1}, textual information \cite{cold_start_review}, knowledge graphs \cite{cold_start_knowledge}, social networks \cite{cold_start_social} and so on. By doing this, representations of users and items are enhanced with additional semantic information so that the demand for interaction data is weakened to some extent. Besides that, the cold-start recommendation could be treated as an application in few-shot learning where only a small number of samples are observed in each task. Similarly, recommendation tasks for new users or items with sparse interactions are naturally divided into meta-training tasks, and meta-learning techniques are widely utilized to alleviate the data insufficiency of cold-start recommendation tasks \cite{MeLU, MAMO}.

\textbf{Click-through-rate prediction.} In online advertising applications, click-through rate (CTR) is a key index to determine the values of published ads \cite{WideDeep, CTR_3, CTR_DIN}. A rational ad auction mechanism should spend more cost on ads with higher CTRs, so as to ensure greater benefits. Therefore,  accurate CTR prediction provided by advertisement publishers could assist investors with subsequent resource allocations. To estimate the click probability of a user-ad pair, recent CTR prediction models usually follow a general framework consisting of two parts including an embedding layer and a prediction layer \cite{WideDeep}. Specifically, the embedding layer first learns latent embedding vectors for both ad/user ids and other rich features. Then the prediction layer is utilized to model feature interaction or dependencies with sophisticated models which are usually well designed as deep neural structures. Despite its success in both academia and industry, the majority of these methods work poorly on new ads due to the lack of embedding learning \cite{Meta-Embedding}. Known as the cold-start problem in CTR prediction, embeddings (especially identity embeddings) of new ads which have limited click records, are hard to be trained as well as other existing ads. As we investigated, meta-learning methods have been studied to strengthen the embedding learning for cold-start ads.

\textbf{Online  recommendation.} In practical large-scale recommender systems, real-time user interaction data are generated and collected continuously. It is necessary to timely refresh the recommendation models previously learned so that dynamic user preferences and trends could be captured \cite{online_1, online_2}. Instead of offline training a model purely based on historical logs, online recommendation attempts to continuously update current recommendation models based on newly arrived data in an online fashion. Online learning strategies and model retraining mechanisms are explored in this field to meet the needs. Due to practical requirements in real-world applications, computation efficiency is a critical factor that should be emphasized. For instance, full retraining over both historical and new samples is an ideal strategy for model refreshing but is pretty impractical for unacceptable time cost \cite{SML}. Therefore, to improve the ability of fast learning, meta-learning has been introduced into online recommendation scenarios and used to quickly capture dynamic preference trends from real-time user interaction data \cite{SML, ASMG}.

\textbf{Point-of-interest recommendation} With the emergence of location-based social networks (LBSNs),  users are willing to share their visited point-of-interests (POIs) through check-in records. LBSN services are supposed to provide personalized recommendations on other POIs that users have not visited. Compared with general item (e.g., product,  music, and movie) recommendation, POI recommendation relies more on discovering spatial-temporal dependencies from historical check-in data. This phenomenon is also very intuitive since users' activities are largely influenced by geospatial and temporal constraints. By incorporating geographical and time information of check-in data, a series of approaches involving spatio-temporal modeling are proposed for POI recommendation \cite{POI_1, POI_2}. Despite their success, the data sparsity issue is obvious in this recommendation scenario since users must arrive at the location of shared POIs. In other words, it is common that users just visited a small number of POIs because of the high cost of data generation. Therefore, meta-learning based POI recommendation methods have been studied to face severe data sparsity \cite{MFNP, Meta-SKR}.

\textbf{Sequential recommendation} The heart of sequential recommendation is to capture evolving user preferences from users' interaction sequences. Different from traditional collaborative filtering methods which organize interactions in the form of user-item pairs, sequential recommendation methods mainly utilize the sequence of previously interacted items of a user as input, and make efforts to discover sequential patterns of user interest evolution. Specifically, representative sequential modeling methods including Markov Chains \cite{MC_1, MC_2}, recurrent neural networks \cite{RNN_1, RNN_2}, and self-attention based networks \cite{SAN_1, SAN_2}, have achieved promising performance in modeling both short-term and long-term interests based on interaction sequences. However, the performances of sequential recommenders usually rely on sufficient items in the sequences. When the number of historical interactions is relatively small, model performance tends to degrade significantly and fluctuate greatly. Consequently, the data sparsity issue also brings stubborn obstacles in the sequential recommendation scenario.

\begin{figure}[tb]
	\centering
	\includegraphics[width = 0.95\textwidth]{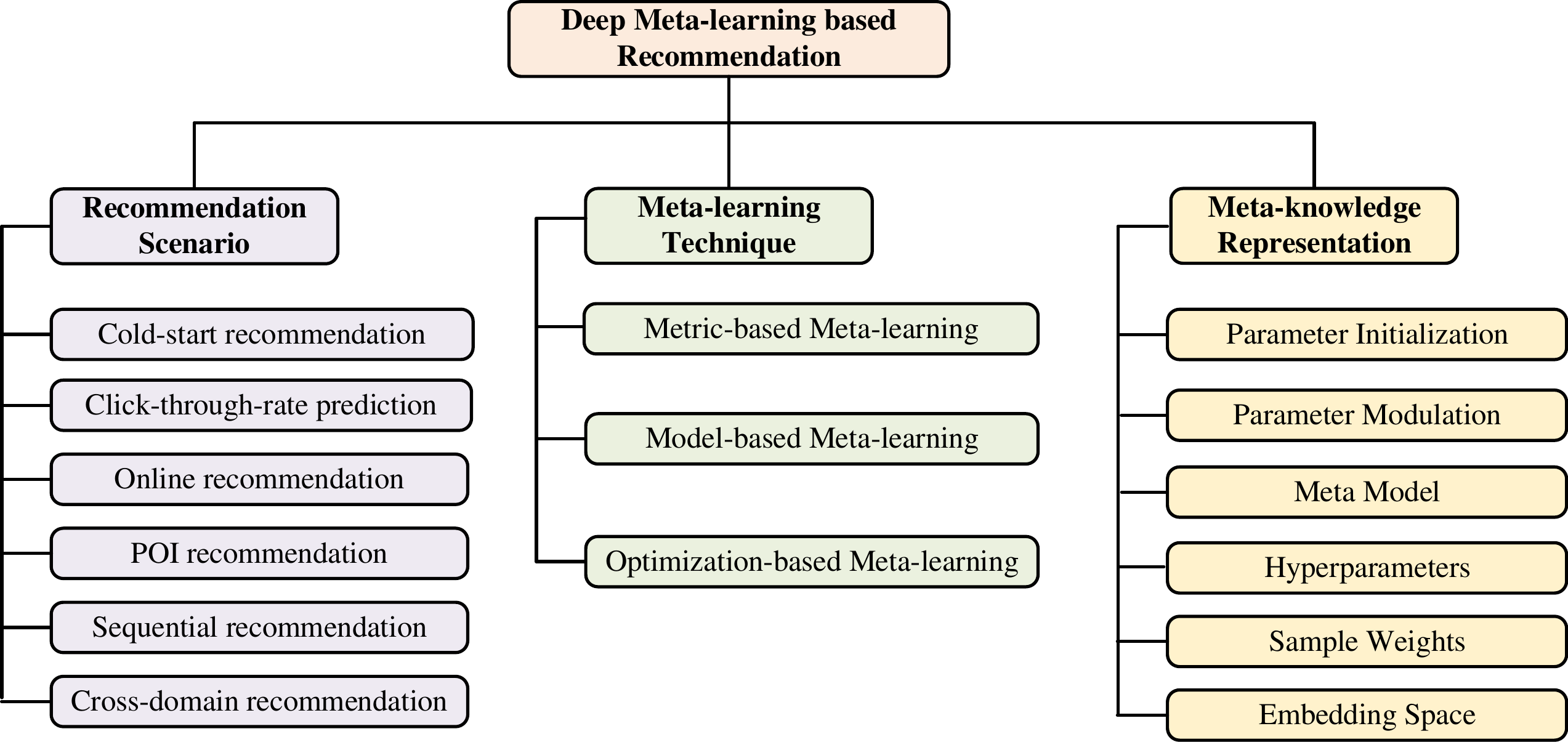}
	\caption{Taxonomy of deep meta-learning based recommendation systems.}
	\label{img1}
\end{figure}

\section{Taxonomy}
\label{sec:3}
In this section, we establish our taxonomy of deep meta-learning based recommendation systems and summarize the characteristics of existing methods according to the taxonomy.

In general, we define our taxonomy in terms of three independent axes, including recommendation scenarios, meta-learning techniques, and meta-knowledge representation. Fig.1 shows the taxonomy. The previous taxonomy of general meta-learning methods proposed in \cite{DML_survey, ML_survey_1} cares more about three categories of meta-learning frameworks as introduced in section 2.1 but pays limited attention to practical applications of meta-learning techniques. In addition, \cite{ML_survey_2} propose a new taxonomy involving three perspectives including meta-representation, meta-optimizer and meta-objective. They provide a more comprehensive breakdown that can orient the development of new meta-learning methods. However, it focuses on the whole meta-learning landscape and is inappropriate to reflect the current research status and application scenarios in deep meta-learning based recommendation systems. Therefore, we concentrate on the recommendation system community and summarize the characteristics of existing works following three dimensions: 

\textbf{Recommendation scenarios (Where)}: This axis presents the specific scenario \emph{where} the meta-learning based recommendation methods are proposed and applied. As introduced in section 2.2, we summarize typical recommendation scenarios into the following groups 1) cold-start recommendation, 2) click-through-rate prediction, 3) online recommendation, 4) point of interest recommendation, 5) sequential recommendation, and 6) others. For clarity, we do not display all involved recommendation scenarios one by one but include less studied scenarios together and denote them as \emph{others}. 

\textbf{Meta-learning techniques (How)}: This axis presents the way how to apply meta-learning to enhance generalization ability over new recommendation tasks. Following the taxonomy in \cite{DML_survey, ML_survey_1}, we also divide meta-learning techniques into three categories including metric-based meta-learning, model-based meta-learning, and optimization-based meta-learning. 

\begin{sidewaystable}[thp]
	\belowcaptionskip = 150mm
	\caption{Summarization of all meta-learning based recommendation methods. We organize  all these methods from hierarchical perspectives of scenarios and meta-learning techniques. We use the following abbreviations. \textbf{Optimi.}: Optimization-based. \textbf{Model}: Model-based. \textbf{Para. Init.}: Parameter Initialization. \textbf{Para. Modu.}: Parameter Modulation. \textbf{Hyperpara.}: Hyperparameter. \textbf{Embedd. Space.}: Embedding Space.}\label{tab2}
	\centering
	\setlength{\tabcolsep}{0.8mm}
	\begin{tabular}{c|ccc|ccc|cccccc}
		\toprule
		\multirow{3}{*}{\textbf{Scenario}} & \multirow{3}{*}{\textbf{Method}} & \
		\multirow{3}{*}{\textbf{Venue}} & \multirow{3}{*}{\textbf{Year}}& \multicolumn{3}{c}{\textbf{Meta-learning Technique}} & \multicolumn{6}{c}{\textbf{Meta-learning Representions}} \\
		\cline{5-13}
		& & & & \textbf{Optimi.} &\textbf{Model} & \textbf{Metric} & \textbf{\makecell[c]{Para.  \\ Init.}} & \textbf{\makecell[c]{Para.\\ Modu.}}  & \textbf{\makecell[c]{Hyper- \\ para. }} & \textbf{\makecell[c]{Meta-\\ Model}} &  \textbf{\makecell[c]{Embed.\\ space}} & \textbf{\makecell[c]{Sample\\ Weight}} \\
		\cline{1-13}
		\multirow{16}{*}{\makecell[c]{ Cold-start \\ Recommendation}} 
		& LWA \cite{LWA} & NIPS & 2017 & & \checkmark & & & \checkmark & & & & \\
		& MeLU \cite{MeLU} & KDD & 2019 & \checkmark  & & & \checkmark &  & & & & \\
		& MetaCS \cite{MetaCS} & IJCNN & 2019 & \checkmark & & & \checkmark &  & \checkmark  & & & \\
		& MetaHIN \cite{MetaHIN} & KDD & 2020 & \checkmark & & & \checkmark & & & \checkmark  & & \\
		& MAMO \cite{MAMO} & KDD & 2020 & \checkmark & & & \checkmark & & & \checkmark & & \\
		& MetaCF\cite{MetaCF} & ICDM & 2020 & \checkmark & &  & \checkmark &  & \checkmark & & &\\
		& TaNP \cite{TaNP} & WWW & 2021 & & \checkmark & & & \checkmark &  & \checkmark &  & \\
		& PALRML \cite{PALRML} & AAAI & 2021 &  \checkmark & & & \checkmark &  & \checkmark  & & &\\
		& MIRec \cite{MIRec} & WWW & 2021 &   & \checkmark & &  & \checkmark & & \checkmark & &\\
		& MPML \cite{MPML} & ECIR & 2021 & \checkmark & & & \checkmark &  & & & & \\
		& PAML\cite{PAML} & IJCAI & 2021 & \checkmark & & & \checkmark &  & & \checkmark & &\\
		& CMML \cite{CMML} & CIKM & 2021 & & \checkmark & &  & \checkmark & & \checkmark & & \\
		& Heater \cite{Heater} & SIGIR & 2021 &   & \checkmark & &  & \checkmark & & \checkmark &  & \\
		& PreTraining \cite{PRE-TRAINING} & SIGIR & 2021 &  &  & \checkmark &  & & &  & \checkmark & \\
		& ProtoCF \cite{ProtoCF} & Recsys & 2021 &  & & \checkmark  & & & & & \checkmark&  \\
		& MetaEDL \cite{MetaEDL} & ICDM & 2021 & \checkmark & &   & \checkmark  & & & & &  \\
		& DML \cite{DML} & AAAI & 2022 & \checkmark & &   & \checkmark  & & & & &  \\
		& PNMTA \cite{PNMTA} & WWW & 2022 & \checkmark & &   & \checkmark  & & & \checkmark & &  \\
		\hline
		\multirow{6}{*}{\makecell[c]{ Click Through \\ Rate Prediction}} 
		& Meta-Embed. \cite{Meta-Embedding} & SIGIR & 2019 & \checkmark & & & \checkmark &  & & \checkmark & & \\
		& TDAML \cite{TDAML} & ACMMM & 2020 & \checkmark & & & \checkmark &  &  & \checkmark  &  & \checkmark\\
		& MWUF \cite{MWUF} & SIGIR & 2021 &  & \checkmark & &  & \checkmark  & & \checkmark & & \\
		& DisNet \cite{DisNet} & Complexity & 2021 & \checkmark & & & \checkmark &  & & \checkmark   & & \\
		& GME \cite{GME} & SIGIR & 2021  & \checkmark & & & \checkmark &  & & \checkmark  & & \\
		& Meta-SSIN \cite{Meta-SSIN} & SIGIR(short) & 2021 & \checkmark  & & & \checkmark &  & & & & \\
		\hline
		\multirow{5}{*}{\makecell[c]{ Point of Interest \\ Recommendation}} 
		& PREMERE \cite{PREMERE} & AAAI & 2021 &  & \checkmark & & &  & & \checkmark  &  & \checkmark \\
		& MetaODE \cite{MetaODE} & MDM & 2021  & \checkmark  & & & \checkmark &  & &  &  &\\
		& MFNP \cite{MFNP} & IJCAI & 2021 & \checkmark &  & & \checkmark &  & &   &  &\\
		& CHAML \cite{CHAML} & KDD & 2021 & \checkmark &  & & \checkmark  &  & & & & \checkmark\\
		& Meta-SKR \cite{Meta-SKR} & TOIS & 2022 & \checkmark &  & &  \checkmark &  & & \checkmark  & &\\
		\bottomrule
	\end{tabular} 
\end{sidewaystable}

\begin{sidewaystable}[htp]
	\belowcaptionskip = 150mm
	\caption{Summarization of all meta-learning based recommendation methods. We organize  all these methods from hierarchical perspectives of scenarios and meta-learning techniques. We use the following abbreviations. \textbf{Optimi.}: Optimization-based. \textbf{Model}: Model-based. \textbf{Para. Init.}: Parameter Initialization. \textbf{Para. Modu.}: Parameter Modulation. \textbf{Hyperpara.}: Hyperparameter. \textbf{Embedd. Space.}: Embedding Space.}\label{tab3}
	\centering
	\setlength{\tabcolsep}{0.5mm}
	\begin{tabular}{c|ccc|ccc|cccccc}
		\toprule
		\multirow{3}{*}{\textbf{Scenario}} & \multirow{3}{*}{\textbf{Method}} & \
		\multirow{3}{*}{\textbf{Venue}} & \multirow{3}{*}{\textbf{Year}}& \multicolumn{3}{c}{\textbf{Meta-learning Technique}} & \multicolumn{6}{c}{\textbf{Meta-learning Representions}} \\
		\cline{5-13}
		& & & & \textbf{Optimi.} &\textbf{Model} & \textbf{Metric} & \textbf{\makecell[c]{Para.  \\ Init.}} & \textbf{\makecell[c]{Para.\\ Modu.}}  & \textbf{\makecell[c]{Hyper- \\ para. }} & \textbf{\makecell[c]{Meta-\\ Model}} &  \textbf{\makecell[c]{Embed.\\ space}} & \textbf{\makecell[c]{Sample\\ Weight}} \\
		\cline{1-13}
		\multirow{7}{*}{\makecell[c]{ Online \\ Recommendation}} 
		& S2Meta \cite{S2Meta} & KDD & 2019 & \checkmark  & & & \checkmark &  & \checkmark  & \checkmark & & \\
		& SML \cite{SML} & SIGIR & 2020 & & \checkmark & &  & \checkmark & & \checkmark & & \\
		& FLIP \cite{FLIP} & IJCAI & 2020 & \checkmark & & & \checkmark &  & & & & \\
		& FORM \cite{FORM} & SIGIR & 2021 & \checkmark  & &  & \checkmark &  & \checkmark &  & &  \\
		& LSTTM \cite{LSTTM} & WSDM & 2022 & \checkmark & & & \checkmark &  & & & & \\
		& ASMG \cite{ASMG} & Recsys & 2021  & & \checkmark & &  & \checkmark  & & \checkmark & &  \\
		& MeLON \cite{MeLON} & AAAI & 2022  & & \checkmark & &  & & \checkmark & \checkmark & &  \\
		\hline
		\multirow{4}{*}{\makecell[c]{ Sequential \\ Recommendation}} 
		& Mecos \cite{Mecos} & AAAI & 2021 & &  & \checkmark & & & & & \checkmark  & \\
		& MetaTL \cite{MetaTL} & SIGIR(short) & 2021 & \checkmark  & & & \checkmark & & & & & \\
		& CBML \cite{CBML} & CIKM & 2021 & \checkmark   & & & \checkmark  & & & \checkmark  & & \\
		& metaCSR \cite{metaCSR} & TOIS & 2022 & \checkmark  & & & \checkmark  & & & &  &
		\\ 
		\hline
		\multirow{2}{*}{\makecell[c]{ Cross Domain \\ Recommendation}} 
		& TMCDR \cite{TMCDR} & SIGIR(short) & 2021 & \checkmark & & & \checkmark &  & & & &\\
		& PTUPCDR \cite{PTUPCDR} & WSDM & 2022 & & \checkmark  & & &  \checkmark & & \checkmark & &\\
		\hline
		\multirow{2}{*}{\makecell[c]{ Multi-behavior \\ Recommendation}} 
		& CML \cite{CML} & WSDM & 2022 & & \checkmark & & & & \checkmark & \checkmark &  &\\
		& MB-GMN \cite{MB-GMN} & SIGIR & 2021 & & \checkmark & & & \checkmark & & \checkmark & &\\
		\hline
		\multirow{6}{*}{\makecell[c]{ Others }} 
		
		& MetaKG \cite{MetaKG} & TKDE & 2022 & \checkmark & & & \checkmark& & & & & \\
		& MetaSelector \cite{MetaSelector} & WWW & 2020 & \checkmark & & & \checkmark & & \checkmark  &  & &\\
		& Meta-SF \cite{Meta-SF} & SDM & 2019 & \checkmark & & &  & & & \checkmark & & \\
		& MetaMF \cite{MetaMF} & SIGIR & 2020 & \checkmark & & & & \checkmark & & \checkmark & & \\
		& MetaHeac \cite{MetaHeac} & KDD & 2021 &\checkmark & & & \checkmark & & & & &\\
		& NICF \cite{NICF} & SIGIR & 2021 &  & \checkmark & & & &  & \checkmark & &\\
		\bottomrule
	\end{tabular} 
\end{sidewaystable}

\textbf{Meta-knowledge representations  (What)}: This axis presents the form of meta-knowledge to be represented so that it could be beneficial for improving the fast learning of recommendation models. After distilling from existing works, we summarize common representations of meta-knowledge as parameter initialization, parameter modulation, hyperparameters, sample weights, embedding space, and meta model. Generally speaking, different meta-learning techniques have distinct characteristics of meta-knowledge representation. For example, parameter initialization is usually achieved under the optimization-based meta-learning while parameter modulation is more likely to belong to model-based meta-learning. However, there are also situations where multiple types of meta-knowledge representations are learned simultaneously in a hybrid manner.

By investigating existing works from the three independent dimensions above, our taxonomy is expected to be able to provide a clear design space for deep meta-learning based recommendation methods. we organize papers according to recommendation scenarios and present characteristics of these works along with the taxonomy in table \ref{tab2} and \ref{tab3}, which lists detailed publication information, and highlights major meta-learning techniques and the forms of meta-knowledge representations. 

\section{Meta-learning Task Construction for Recommendation}
\label{sec:4}

In this section, we summarize different ways of meta-learning recommendation task construction used in the literature. As discussed in section 2.1, one major difference between the meta-learning paradigm and the regular deep learning paradigm is the setup of task division. We will first introduce the general form of constructing meta-learning tasks and then present practical ways adopted in deep meta-learning based recommendation methods, which are quite different from other fields.

In general, meta-learning methods usually follow the setting of constructing disjoint meta-training tasks $\mathcal{D}^{train}$ and meta-test tasks $\mathcal{D}^{test}$. Each task is split into a set of training instances (named support set $\mathcal{S}_{i}$) and a disjoint set of evaluation instances (named query set $\mathcal{Q}_{i}$). The objective of each task is to learn quickly from the support set $\mathcal{S}_{i}$, so as to perform better over \textbf{unseen instances} in the query set $\mathcal{Q}_{i}$. At a single task level, its learning objective is similar to the regular deep learning paradigm, except that data insufficiency of the task is usually emphasized in meta-learning. When considering the whole task distribution (or multiple tasks), a higher level of learning objective (i.e., meta-optimization objective) is defined as better performance on evaluation instances of \textbf{unseen tasks} (i.e., $\mathcal{D}^{test}$). Consequently, the setup of task division above is consistent with the meta-optimization objective, facilitating the  evaluation of generalization ability and the fast learning ability of meta-learning methods over multiple new tasks. 

Different from meta-learning task construction settings in other application domains, constructing meta-learning recommendation tasks should meet the special needs of different recommendation scenarios. For popular few-shot classification tasks such as image recognition and objective detection, a commonly used setting is $N$-way, $K$-shot classification \cite{MAML}. Specifically, based on a pool with a large number of classes, a task is obtained by randomly sampling $N$ classes first and then sampling K instances belonging to each class. The $K$ is usually set as a small number to meet the requirements of a few-shot task. For meta-learning tasks in natural language processing, Lee et al. \cite{NLP_meta_survey_1} summarize different settings of task construction including cross-domain, cross-lingual, cross-problem, domain-generalization, and homogenous task augmentation. For instance, tasks in the cross-domain setting are from different domains (e.g., texts from news and laws are considered as different domains), while tasks in the cross-lingual setting are divided based on different languages. Overall, The settings of meta-learning tasks in the other fields mentioned above are closely related to the task objectives and data characteristics. Therefore, we specially discuss the construction of meta-learning recommendation tasks and present how existing meta-learning methods perform task division with interaction data from recommendation systems.

According to common properties belonging to interactions in a task, we mainly summarize the task construction ways into four categories, including \emph{user-specific} task, \emph{item-specific}, \emph{time-specific} task and \emph{sequence-specific} task. To be mentioned, there are a few works that have tried other ways but the number is relatively small. We organize them all in the category named $others$. Table \ref{tab4} shows the summary of works adapting each category of task construction.
\begin{table}[t]
	\caption{Summary of task construction in meta-learning based recommendation methods.}\label{tab4}
	\centering
	\setlength{\tabcolsep}{1mm}
	\begin{tabular}{c|c}
		\toprule
		\textbf{Task  Construcution} & \textbf{Methods} \\
		\cline{1-2}
		\textbf{User-specific} & \makecell[c]{LWA \cite{LWA},MeLU \cite{LWA},MetaCS \cite{MetaCS}, MetaHIN \cite{MetaHIN},  MAMO \cite{MAMO} \\ TaNP \cite{TaNP}  PALRML \cite{PALRML}, MPML \cite{MPML},   PAML \cite{PAML}, CMML \cite{CMML}, \\  Heater \cite{Heater}, PNMTA \cite{PNMTA}, 
			Meta-SSIN \cite{Meta-SSIN},  MFNP \cite{MFNP},\\  FORM \cite{FORM}, PTUPCDR \cite{PTUPCDR}, MetaKG \cite{MetaKG}, MetaEDL \cite{MetaEDL}} \\
		\cline{1-2} 
		\textbf{Item-specific} & \makecell[c]{MIRec \cite{MIRec}, ProtoCF \cite{ProtoCF},  Meta-Embed. \cite{Meta-Embedding},  TDAML \cite{TDAML}, \\ MWUF \cite{MWUF}, DisNet \cite{DisNet}, GME \cite{GME}, Mecos \cite{Mecos}}  \\
		\cline{1-2}
		\textbf{Time-specific} & \makecell[c]{DML \cite{DML}, SML \cite{SML}, LSTTM \cite{LSTTM}, ASMG \cite{ASMG}, MeLON \cite{MeLON}} \\
		\cline{1-2}
		\textbf{Sequence-specific} & \makecell[c]{FLIP \cite{FLIP}, Meta-SKR \cite{Meta-SKR}, MetaTL \cite{MetaTL}, CBML \cite{CBML}, MetaCSR \cite{metaCSR}} \\
		\cline{1-2}
		\textbf{Others} & \makecell[c]{PreTraining \cite{PRE-TRAINING}, PREMERE \cite{PREMERE}, MetaODE \cite{MetaODE}, CHAML \cite{CHAML}, \\
			S2Meta \cite{S2Meta}, TMCDR \cite{TMCDR} } \\
		\bottomrule
	\end{tabular} 
\end{table}

\begin{figure}[tb]
	\centering
	\includegraphics[width = 0.65\textwidth]{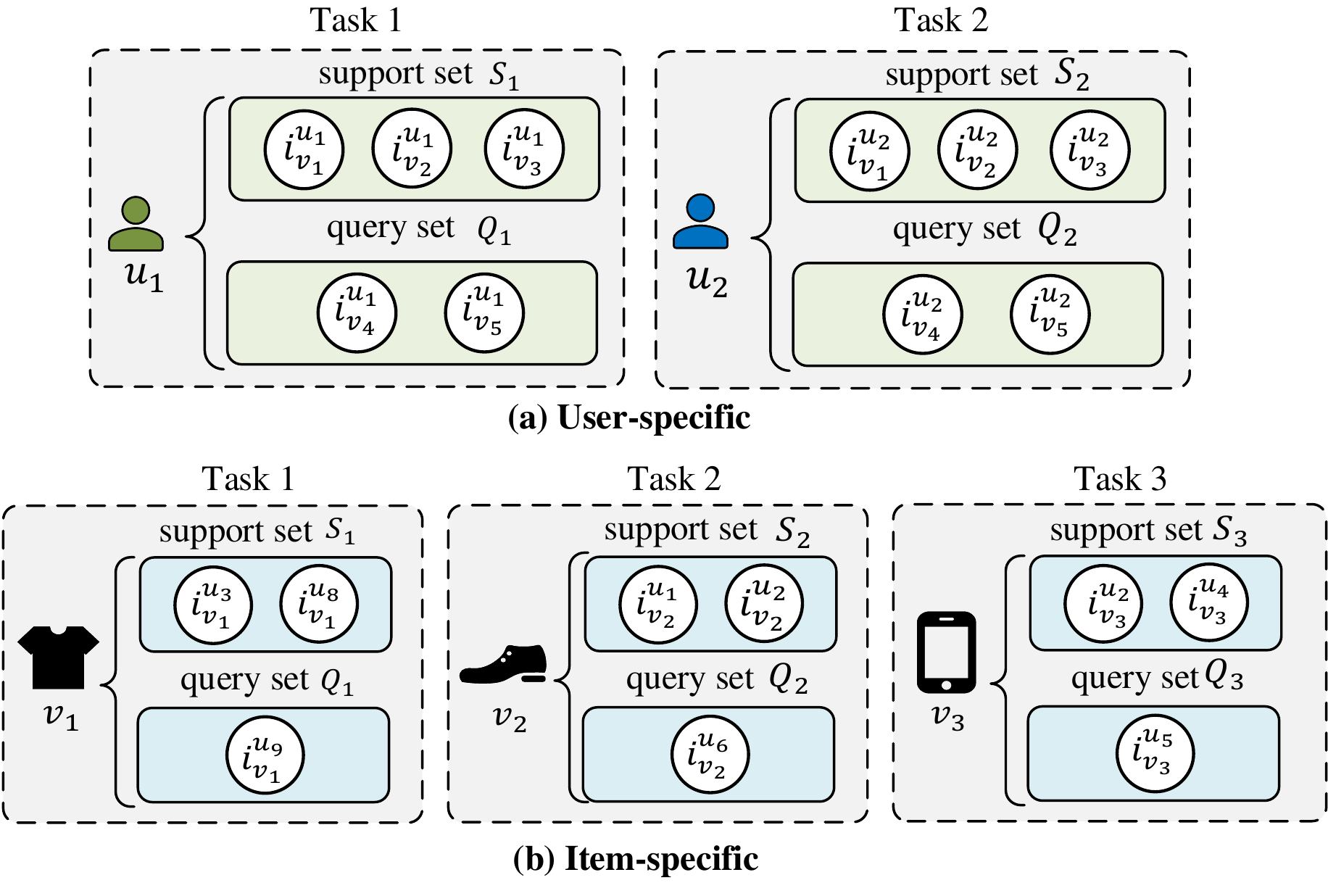}
	\caption{Illustration of task construction for user-specific tasks and item-specific tasks.}
	\label{img2}
\end{figure}

\textbf{User-specific Task.} As observed in Table 4, the most typical way of task construction is based on users. Since the user cold-start issue is the most long-standing problem in recommendation systems, quickly learning preferences from users' limited interactions is a critical task to be solved. In the setting of user-specific task $\mathcal{T}_{i}$, all instances of a task including both the support set $\mathcal{S}_{i}$ and the query set $\mathcal{Q}_{i}$ are belonging to the same user. Learning preferences of different users are naturally treated as different tasks. Give a illustrative example shown in Fig \ref{img2} (a). For a user-specific task of a specific user $u_1$, all his interactions are split into a support set $\mathcal{S}_{1} = \{(v_{j},i^{u_1}_{v_j})\}_{j=1}^{3}$ and a query set $\mathcal{Q}_{1} = \{(v_{j},r^{u_1}_{v_j})\}_{j=4}^{5}$, where $i^{u_1}_{v_j}$ could be a explicit rating score or a implicit feedback between user $u_1$ and item ${v_j}$. The goal of each user-specific task is to train a model on the support set and evaluation on the interactions in the query set of the same user. From the perspective of the meta-optimization objective, meta-learning methods are expected to extract meta-knowledge about user preference learning from a sufficient number of user-specific tasks $\mathcal{D}^{train}$. Then when faced with unseen user-specific tasks from new users, the meta-knowledge should work as prior experiences to facilitate preference learning. 

\textbf{Item-specific Task.} Symmetric with the user-specific task, an item-specific task is constructed based on all instances involving the same item. From the view of an item, interaction instances are grouped based on different items. As illustrated in Fig \ref{img2} (b), three item-specific tasks are constructed according to three different items including a shirt, a shoe, and a phone.
Similar to user-specific tasks, meta-learning based item-specific tasks usually aim at tackling the item cold-start problem. In this setting, the support set and the query set of a task cover all interactions between multiple users and the same item. The goal of each item-specific task is to predict the ratings or interaction probabilities of evaluation instances in the query set after observing the support set. By extracting meta-knowledge across multiple item-specific tasks, meta-learning methods could quickly perceive the overall preference for cold-start items, making accurate predictions and recommendations.

\begin{figure}[tb]
	\centering
	\includegraphics[width = 0.80\textwidth]{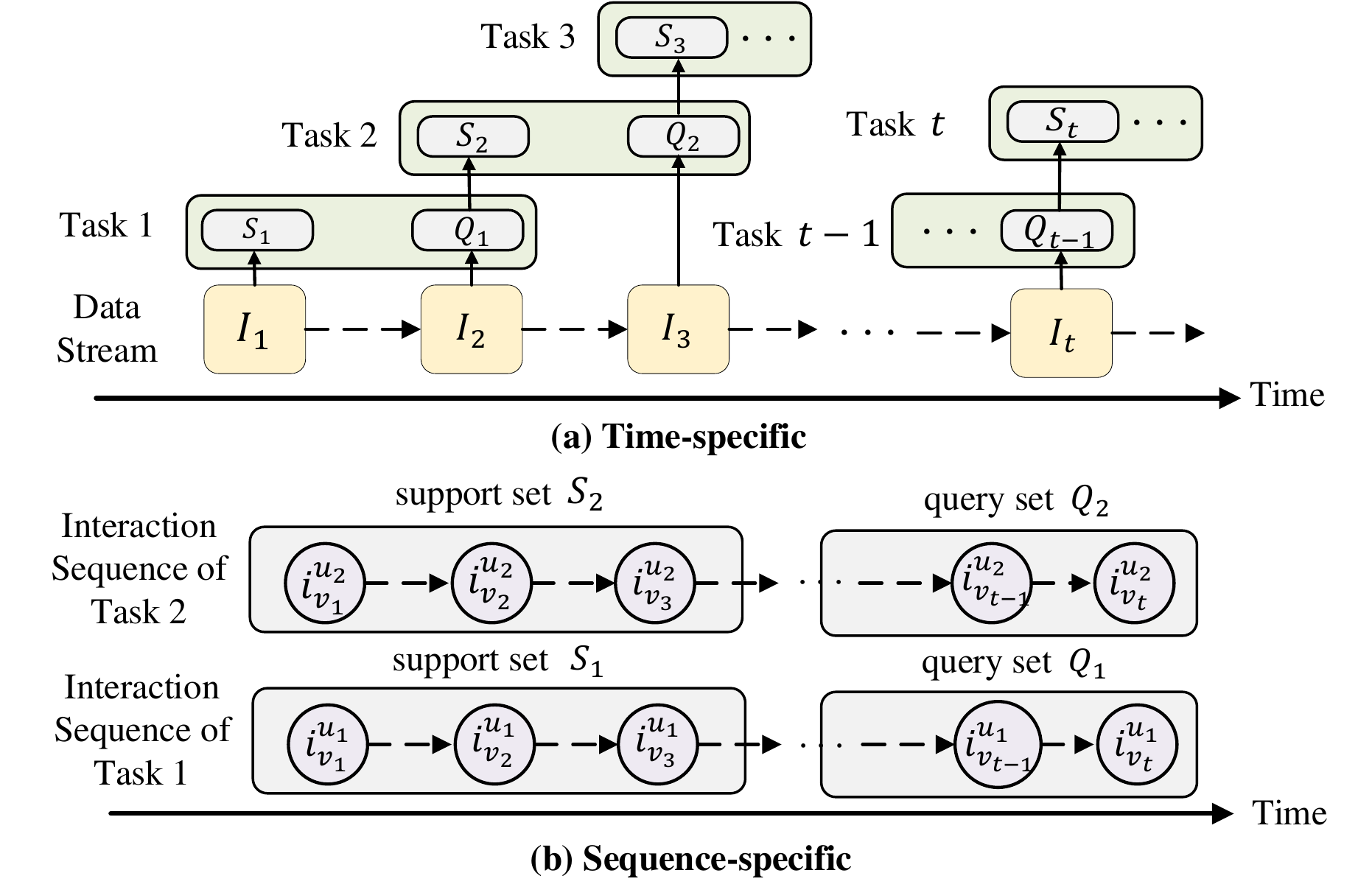}
	\caption{Illustration of task construction for time-specific tasks and sequence-specific tasks.}
	\label{img3}
\end{figure}

\textbf{Time-specific Task.} In this setting, interaction data in recommendation systems are split into different tasks according to different time slots. Specifically, interaction data are considered as collected continually and arrived in the form of data streaming. Formally, at the time $t$, data currently collected is denoted as $I_i = \{(u_i,v_j,i^{u_i}_{v_j})\}^{M}$. Different from user-specific or item-specific settings, interactions in time-specific tasks are no longer distinguished by users or items. As shown in Fig \ref{img3} (a), time-specific tasks are sequentially constructed with data in two successive time slots. For instance, for the task at time $2$, the support set consists of the data block $I_2$, i.e., data collected at the current time slot. For the query set, data block $I_2$ in the next time slot 3 is utilized as evaluation data. The reason for this setting is that the goal of a time-specific task is usually to efficiently update models in an online setting so that the updated model could still perform well in the next period. Meta-learning can also be used to facilitate the efficiency of model online updates by gradually extracting meta-knowledge from sequential time-specific tasks.

\textbf{Sequence-specific Task.} As illustrated in Fig \ref{img3} (b), sequence-specific tasks are also constructed with temporal information considered. Different from time-specific tasks which collect data at the system level, the sequence-specific setting treats interaction sequences of different users or different sessions as different tasks. For example, the whole interaction sequence of user $u_1$ is denoted as $\{(v_{1},i^{u_1}_{v_1}),(v_{2},i^{u_1}_{v_3}),...,(v_{t},i^{u_1}_{v_t})\}$ which is ordered by interaction timestamps. For constructing a sequence-specific task, the interaction sequence with the length $t$ is usually split into two parts. The former $K$ interactions are allocated as the support set, while the latter $t-K$  interactions are allocated as the query set. There are two major differences between user-specific tasks and sequence-specific tasks. First, sequence-specific tasks are not restricted by integrating interaction users' history.  Anonymous sessions can also be independent interaction sequences. Second, the form of instances in sequence-specific tasks are usually subsequences of the whole interaction sequence, while user-specific tasks have interaction pairs.

\textbf{Others.} Besides the four types of tasks mentioned above, several works also explore other ways of task construction. Scenario-specific tasks \cite{S2Meta} are divided according to different scenarios (e.g., tags, themes, or categories of items) in recommendation systems. Special for POI recommendation,  city-specific tasks \cite{MetaODE, CHAML} organize interactions according to different cities, so that meta-knowledge could be extracted across multiple city-specific tasks and benefits data-sparse cities. Different from user-specific tasks which utilize interactions of a single user as a task, interactions of multiple users could also be combined and treated as one task \cite{TMCDR, PREMERE}. Specifically,  in cross-domain recommendation systems,  Zhu et al. \cite{TMCDR} randomly sample two groups of overlapping users (denoted as $U_a$ and $U_b$) and construct a cross-domain meta-learning task by gathering all interactions of multiple users as a support set (i.e., $S_i = D_a$) and a query set (i.e., $Q_i = D_b$), respectively. The goal of each task is to learn an embedding mapping model from a source domain to a target domain for better performance over cold-start users in the target domain (simulated with $D_b$), while meta-learning contributes to the learning of the mapping model across multiple tasks. With a similar strategy of task construction, Kim et al. \cite{PREMERE} also separately samples two groups of multiple users as training data in two update phases of a meta-learning task. Besides recommendation tasks, Hao et al. \cite{PRE-TRAINING} construct reconstruction tasks as pretraining tasks in their proposed meta-learning based cold-start recommendation method. Each reconstruction task consists of a target user and $K$ samples neighboring users and aims to reconstruct the target user's embedding with his neighbors.

\section{Meta-leanring Methods for Recommendation Systems }
\label{sec:5}
In this section, we look in more detail at meta-learning based recommendation methods in the literature. In general, we introduce how meta-learning methods facilitate the progress of recommendation systems in different recommendation scenarios. In each recommendation scenario, we summarize characteristics of related works and discuss methods about their ways of applying meta-learning. 

\subsection{Meta-learning in Cold-start Recommendation} 




In cold-start recommendation scenarios, users who conduct a small number of interactions or items which are involved in few interactions are emphasized when making recommendations, so as to boost the overall performance of recommendation systems. As commonly known, few-shot learning is the most common application of meta-learning. In recommendation systems, as an analogy to the few-shot learning problem, cold-start recommendation is also paid more attention and well studied by meta-learning based methods. Here, we summarize how existing works apply meta-learning to alleviate the cold-start issues for both cold-start users and items into different groups, including \emph{optimization-based parameter initialization}, \emph{optimization-based parameterized hyperparameters}, \emph{model-based parameter modulation} and \emph{metric-based embedding space learning}.  Next, we will elaborate on different categories of methods and introduce details of concrete methods.

\begin{figure}[tb]
	\centering
	\includegraphics[width = 0.85\textwidth]{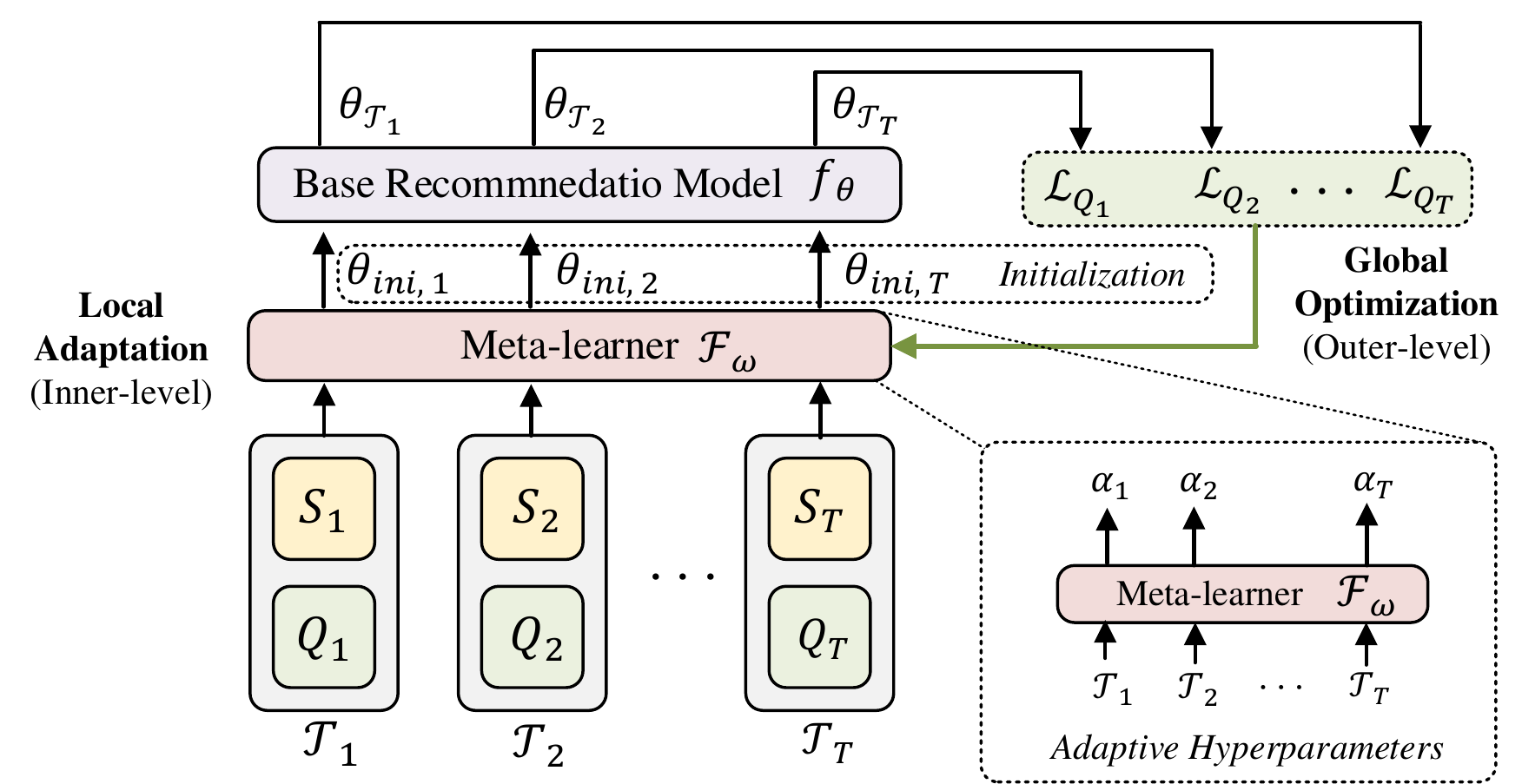}
	\caption{Illustration of the framework of Optimization-based Parameter Initialization and Adaptive Hyperparameters. Based on two levels of optomization including local adaptation and global optimization, the optimization-based meta-learner is updated across meta-training tasks. Both parameter initialization and adaptive hyperparameters could be learned according to different designs of meta-learners. }
	\label{img4}
\end{figure}

\begin{table}[t]
	\caption{Details of recommendation models with optimization-based meta-learning methods in cold-start recommendation.  The key techniques in both inner-level update and outer-level optimization are presented.}\label{tab5}
	\centering
	\begin{tabular}{cccl}
		\toprule
		\textbf{Method} & \textbf{\makecell[c]{Cold-start \\ Object}} &
		\textbf{\makecell[c]{Meta-knowledge \\ Representation}} &  \textbf{\makecell[c]{Key Techniques in \\  Bi-level Optimization}}   \\
		\midrule
		MeLU \cite{MeLU} & User \& Item & Parameter Initialization & \emph{Inner}: FCN \emph{Outer:} MAML
		\\
		\hline
		MetaCS \cite{MetaCS} & User &  \makecell[c]{Parameters Initialization \\ \& Hyperparameter} & \emph{Inner}: FCN   \emph{Outer:} MAML \\
		\hline
		MetaHIN \cite{MetaHIN} & User \& Item &  \makecell[c]{Parameter Initialization \\ \& Meta Model} & \makecell[l]{\emph{Inner}: FCN  \\ \emph{Outer:} MAML + HIN} \\
		\hline
		MAMO \cite{MAMO} & User \& Item  & \makecell[c]{Parameter Initialization \\ \& Meta Model} & \makecell[l]{\emph{Inner}: FCN  \\ \emph{Outer:} MAML + Memories}\\
		\hline
		MetaCF \cite{MetaCF} & User &  \makecell[c]{Parameter Initialization \\ \& Hyperparameter} & \makecell[l]{\emph{Inner}: FISM \cite{FISM} or NGCF \cite{NGCF} \\ \emph{Outer:} MAML} \\
		\hline
		PALRML \cite{PALRML}  & User &  \makecell[c]{Parameter Initialization \\ \& Hyperparameter}  & \makecell[l]{\emph{Inner}: FCN  \\ \emph{Outer:} MAML + Adaptive Learning Rate} \\
		\hline
		MPML \cite{MPML} & User \& Item & Parameter Initialization & \makecell[l]{\emph{Inner}: FCN  \\ \emph{Outer:} MAML + Clustering} \\
		\hline
		PAML\cite{PAML} & User \& Item &  \makecell[c]{Parameter Initialization \\ \& Meta Model} & \makecell[l]{\emph{Inner}: HIN + Social  \\ \emph{Outer:} MAML + Gating} \\
		\hline
		MetaEDL \cite{MetaEDL} &  User &  \makecell[c]{Parameter Initialization} & \emph{Inner}: FCN   \emph{Outer:} MAML \\
		\hline
		DML \cite{DML} & User &  \makecell[c]{Parameter Initialization } & \makecell[l]{\emph{Inner}: FCN + RNN  \\ \emph{Outer:} MAML} \\
		\hline
		PNMTA \cite{PNMTA} & User &  \makecell[c]{Parameter Initialization \\ \& Meta-Model} & \makecell[l]{\emph{Inner}: FCN  \\ \emph{Outer:} MAML + Parameter Modulation} \\
		\bottomrule
	\end{tabular} 
\end{table}

\textbf{Optimization-based Parameter Initialization.} 
Table \ref{tab5} shows the summary of optimization-based meta-learning methods in cold-start recommendation from three perspectives, i.e., cold-start object, meta-knowledge representation, and key techniques used in the bi-level optimization framework. Existing methods generally fall into two categories according to two forms of meta-knowledge representations, including \emph{parameter initialization} and \emph{adaptive hyperparameters}. We present a general framework for both optimization-based parameter initialization and adaptive hyperparameters in Fig \ref{img4}. In the following, We discuss concrete methods for parameter initialization in this part and adaptive hyperparameters in the next part.

The basic idea of \emph{optimization-based parameter initialization} is defining the meta-knowledge $\omega$ as the initial parameters of base recommendation models and then updating the parameter initialization in the form of bi-level optimization. Inspired by the idea of model-agnostic meta-learning\cite{MAML}, Lee et.al \cite{MeLU} firstly introduce the MAML framework to cold-start recommendation and propose \textbf{MeLU}, which aims to learn global parameter initialization of a neural network based recommendation model as prior knowledge. The base model $f_{\theta}$ is implemented using fully connected neural networks (FCNs), which act as a personalized user preference estimation model. Here, $\theta$ include transformation parameters $\bm{W}$ and bias parameters $\bm{b}$ of both hidden layers and the final output layer in the base recommendation model, which are to be initialized with globally learned parameter initialization $\omega$ via $\theta \leftarrow \omega$. Following the bi-level optimization procedure, MeLU constructs user cold-start tasks and locally updates the parameters of the personalized recommendation model for each user $u_i$ as the equation (6). After the local update process, a user-specific recommendation model $f_{\theta_{\mathcal{T}_{i}}}$ is especially learned for the task $\mathcal{T}_{i}$, and employed to make preference predictions on its unseen query set $\mathcal{Q}_{i}$. In the global optimization procedure, global parameter initialization $\theta$, which is applied to the local update processes of multiple meta-training tasks simultaneously, is optimized by minimizing the summed loss on query sets as equation (7). After iterative global update steps during the meta-training phase, the global parameter initialization $\omega$ is supposed to have abilities to quickly adapt to new cold-start recommendation tasks in the meta-testing set $\mathcal{D}^{test}$. In MeLU, the parameters of the user preference estimation model are optimized under the MAML framework while user/item embeddings are only globally updated. In addition, MeLU is evaluated as effective in handling both user and item cold-start issues by dividing both user and item into existing groups and new groups. 

Drawing on the idea of globally learning model initialization parameters across multiple cold-start tasks, some other works are also proposed with the help of the original MAML framework. On the basis of MeLU, Chen et al.\cite{MPML} propose a multi-prior meta-learning approach \textbf{MPML} which equips multiple sets of initialization parameters. For a cold-start task, which set of initialization to be assigned depends on which performs better after local update over its support set. Besides simple FCN-based collaborative filtering models, optimization-based meta-learning also have been utilized to learn initialization for different forms of recommendation models. For instance, \textbf{MetaEDL} \cite{MetaEDL} adopts the MAML framework to learn initialization parameters of an evidential learning enhanced recommendation model which additionally assigns evidence to predicted interactions. Considering the temporal evolution of user preferences, 
\textbf{DML} \cite{DML} is designed to continuously capture time-evolving factors from all historical interactions of a user and fastly learn time-specific factors based on a small number of current interactions.  Specifically, the module for capturing time-specific factors is learned under the MAML framework in order to fastly adapt to each time period where the number of the user's interactions is usually small.

\begin{figure}[tb]
	\centering
	\includegraphics[width = 0.95\textwidth]{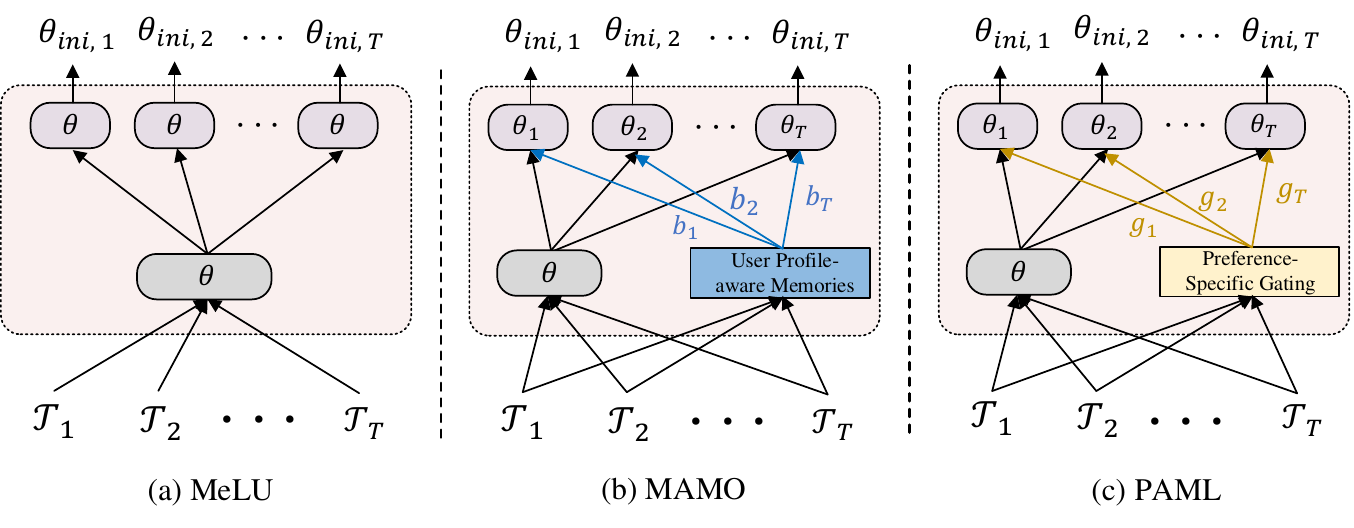}
	\caption{Illustration of different parameter initializationn strategies in three representative methods including \textbf{MeLU}, \textbf{MAMO} and \textbf{PAML}. In short, \textbf{MeLU} shares global initialization among all tasks while \textbf{MAMO} and \textbf{PAML} tailor task-specific initialization  considering user profile and user preferences respectively.}
	\label{img5}
\end{figure}

One promising line of extending the MAML framework is to take the \emph{task heterogeneity} issue into consideration by tailoring task-specific initialization for different tasks \cite{MAMO, PAML, PNMTA}. We present the core idea of initialization strategies in two representative works in Fig \ref{img5}. One representative work \textbf{MAMO} \cite{MAMO} is proposed to provide a personalized bias term when initializing the recommendation model parameters. Specifically, memory networks are introduced into optimization-based meta-learning as external memory units to store task-specific fast weight memories. Before assigning the global initialization learned under the MAML framework to base model, MAMO applies memory units to generate a personalized bias term $b_{u_i}$ and obtain a task-specific initialization $\theta_{u_i} \leftarrow \omega - \tau b_{u_i}$. $b_{u_i}$ is generated by querying fast weights memories $M_W$ with profile representation $p_{u_i}$ of a given user ${u_i}$ as follows: 
\begin{eqnarray} 
b_{u_i} &=& a_{u_i}^T M_W \\
a_{u_i} &=& attention(p_{u_i}, M_P)
\end{eqnarray}
where $M_P$ is the profile memories stored in the training process and $M_W$ is the fast weights memories storing training gradients as fast weights. As for the model and memories optimization, two memory matrices are updated over the training task of ${u_i}$ as follows: 
\begin{eqnarray} 
M_P &=& \lambda (a_{u_i} p_{u_i}^T) + (1-\lambda)M_P \\
M_W &=& \delta (a_{u_i} \nabla_{\theta}  \mathcal{L}(f_{\theta}, \mathcal{S}_{i} )) + (1-\delta) M_W
\end{eqnarray}
where $\lambda$ and $\delta$ are hyperparameters as memory update ratios. Note that we only present one part of the utilization of memories in MAMO, while more details and extensions could be seen in the original paper \cite{MAMO}. Consequently, by injecting the profile-aware initialization bias $b_{u_i}$, MAMO tailors task-specific initialization $\theta_{u_i}$ to copy with task heterogeneity issue w.r.t. user profiles.

Following the same idea of customizing task-specific initialization, Wang et al. \cite{PAML} also argue that similar prior knowledge should be shared by users with similar preferences. Therefore, a preference-adaptive meta-learning approach \textbf{PAML} is proposed to adjust the globally shared prior initialization $\theta$ to the preference-specific initialization $\theta_{u_i}$  by applying an external meta model. Specifically, the meta model acts as a preference-specific adapter by incorporating social relations from social networks and semantic relations from heterogeneous information networks (HINs). When customizing the preference-specific initialization $\theta_{u_i}$, a series of preference-specific gates $\bm{g_{u_i}}$ are designed to control how much prior knowledge is shared, implemented as follows:
\begin{eqnarray}
\bm{g}_{u_i} &=& \sigma(\bm{W}_g\bm{u}_i + \bm{b}_g) \\
\theta_{u_i} &=& \theta \circ \bm{g}_{u_i}
\end{eqnarray}
%
%
where $\bm{u}_i$ is the user preference representation learned from not only interactions of the user as well as representations of his/her explicit friends extracted based on social relations and implicit friends extracted from semantic relations, respectively. Since user relations are comprehensively modeled by incorporating both social networks and HINs, final user preference representation $\bm{u}_i$ is supposed to trigger similar gates for users who share similar preferences. Finally, after obtaining preference-specific initialization $\theta_{u_i}$, optimization-based meta-learning (i.e., MAML framework) is utilized to optimize parameters of both the base recommendation model and the meta model. Here, the base recommendation model includes the preference modeling module previously discussed and an FCN-based rating prediction module. Different from MAMO which focuses on user profile information, PAML distinguishes different tasks mainly based on multiple types of user relations. 

Without incorporating external task relations for revealing differences among tasks, Pang et al. \cite{PNMTA} propose \textbf{PNMTA} to discover implicit task distribution from users' interaction contexts and perform task-adaptive initialization adjustment. Specifically, a meta model $\mathcal{F}_\omega$  is designed to generate task-specific initialization $\theta_{u_i}$ for the base prediction model by conducting parameter modulation as follows: 
\begin{eqnarray}
\bm{w}_i, \bm{b}_i  &=& \mathcal{F}_\omega(\bm{t_i})\\
\theta_{u_i} &=& \bm{w}_i \odot \theta + \bm{b}_i
\end{eqnarray}
where $\bm{t_i}$ is the task vector learned by aggregating all interaction representations. Conditioned on the task representation, the meta model generates task-adaptive modulation signals, i.e., parameters of the modulation function. Here, we present feature-wise linear modulation (FiLM) while other types of modulation functions such as channel-wise modulation and soft attention modulation are also discussed in the original paper. In the meta-training phase, both parameters of the meta-model $\omega$ and global initialization $\theta$  of the base model are optimized under the MAML framework.

Besides the extension over the meta-learning framework, \textbf{MetaHIN} \cite{MetaHIN} is proposed to augment cold-start tasks from the perspective of task construction. Specifically, different from merely regarding interacted items of a user as the support set  $\mathcal{S}_{i}$, MetaHIN incorporates multifaceted semantic contexts $\mathcal{S}_{i}^{\mathcal{P}}$ into tasks based on multiple meta-paths $\mathcal{P} = \{p_1, p_2,...,p_n\}$ of heterogeneous information network (HIN). For each meta-path $p_k$, a set of items that are reachable from user $u_i$ are obtained via $p_k$, denoted as $\mathcal{S}_{i}^{p_k}$. By doing this, the semantic-enhanced support set is obtained as $(\mathcal{S}_{i}, \mathcal{S}_{i}^{\mathcal{P}})$, and semantic-enhanced query set is obtained similarly as $(\mathcal{Q}_{i}, \mathcal{Q}_{i}^{\mathcal{P}})$. After constructing the semantic-enhanced tasks above, a co-adaptation meta-learner is designed to perform both semantic- and task-wise adaptation to enhance the ability of local adaptation for each user.  task-wise adaptation to enhance the ability of local adaptation for each user. The co-adaptation adaptation focuses on adapting to different semantic spaces induced by different meta-paths, respectively. Overall, the conventional local adaptation phase in MAML is first augmented from the data level by constructing semantic-enriched tasks and then enhanced with a co-adaptation meta-learner by designing two levels of local adaptation.

\textbf{Optimization-based Adaptive Hyperparameters.}  
Besides parameter initialization of based recommendation models, several works also leverage meta-learning to learn adaptive hyperparameters for different cold-start tasks. For instance, \textbf{MetaCS} \cite{MetaCS} adopts the similar bi-level optimization procedure as the MeLU, and additionally meta-update the value of local learning rate $\alpha$ when performing global optimization. The updating equation of the local learning rate is as follows: 
\begin{equation} 
\alpha \leftarrow \alpha - \beta \nabla_\alpha \sum\nolimits_{\mathcal{T}_{i} \in \mathcal{D}^{train}} \mathcal{L}(f_{\theta_{\mathcal{T}_{i}}}, \mathcal{Q}_{i} ),
\label{eq-3}
\end{equation}
where $\alpha$ is the parameterized learning rate for the local update and $\beta$ is a fixed learning rate for the global update. They argue that the manually fixed learning rate may make the model unable to converge. In this way, not only model parameters of the base model but also hyperparameters, e.g., learning rates, are meta-learned to provide prior knowledge. To be mentioned, the learnable update ratio here is merely globally optimized but not updated during the local adaptation of each task. 

With collaborative filtering methods as the base model, \textbf{MetaCF} \cite{MetaCF} also leverages MAML framework to meta-learn initialization for learnable parameters such as item embeddings in FISM \cite{FISM} and embedding transformation parameters in NGCF \cite{NGCF}. Similar to MetaCS \cite{MetaCS}, MetaCF also adopts a flexible update strategy by learning appropriate learning rates automatically. While performing task construction, MetaCF adopts another two strategies including dynamic subgraph sampling and potential interactions extraction, which inject dynamicity and semantics into the recommendation tasks.

Similarly, Yu et al. \cite{PALRML} proposes a personalized adaptive learning rate meta-learning approach \textbf{PALRML} which sets different learning rates for different users to find task-adaptive parameters for each task. They argue that assuming uniform user distribution in recommendation systems may lead to the over-fitting problem of major users with similar features. In other words, minor users whose features are different from the major ones may not be focused on. Therefore, PALRML performs user-adaptive learning rate based meta-learning to improve the performance of the basic MAML framework. Specifically, the local adaptation on each task $\mathcal{T}_{i}$ is adjusted as:
\begin{equation} 
\theta_{\mathcal{T}_{i}} = \theta - \alpha(h_i) \nabla_\theta \mathcal{L}(f_\theta, \mathcal{S}_{i} ).
\label{eq-8}
\end{equation}
where $\alpha(h_i)$ is a mapping function for assigning an appropriate learning rate for each user $u_i$ according to the user's feature embedding $h_i$. Three different strategies including adaptive learning rate based, approximated tree-based, and regularizer-based are designed to provide personalized learning rates. Low space complexity and good prediction performance are supposed to be achieved simultaneously. 

\begin{table}[t]
	\caption{Details of recommendation models with model-based meta-learning methods in cold-start recommendation. The key role of designed meta models in different methods is summarized.}\label{tab6}
	\centering
	\begin{tabular}{cccl}
		\toprule
		\textbf{Method} & \textbf{\makecell[c]{Cold-start\\ object}} &  \textbf{\makecell[c]{Base  Model}} & \textbf{\makecell[c]{Key Role of Meta Model}} \\
		\midrule
		LWA \cite{LWA} & Item &  LR / FCN   & Task-dependent Parameter Generation\\
		\hline
		TaNP \cite{TaNP} & User & \makecell[c]{Encoder \& \\ Decoder} & Task Relevance aware Parameter Modification \\
		\hline
		MIRec \cite{MIRec}  & Item &  FCN  & \makecell[l]{Parameter Generation from few-shot models \\ to many-shot models} \\
		\hline
		CMML \cite{CMML} &  User & FCN & Task-dependent Parameter Modification \\
		\hline
		Heater \cite{Heater} &   User \& Item  & FCN  &  Mixture-of-Experts based Parameter Integration \\
		\bottomrule
	\end{tabular} 
\end{table}

\begin{figure}[tb]
	\centering
	\includegraphics[width = 0.95\textwidth]{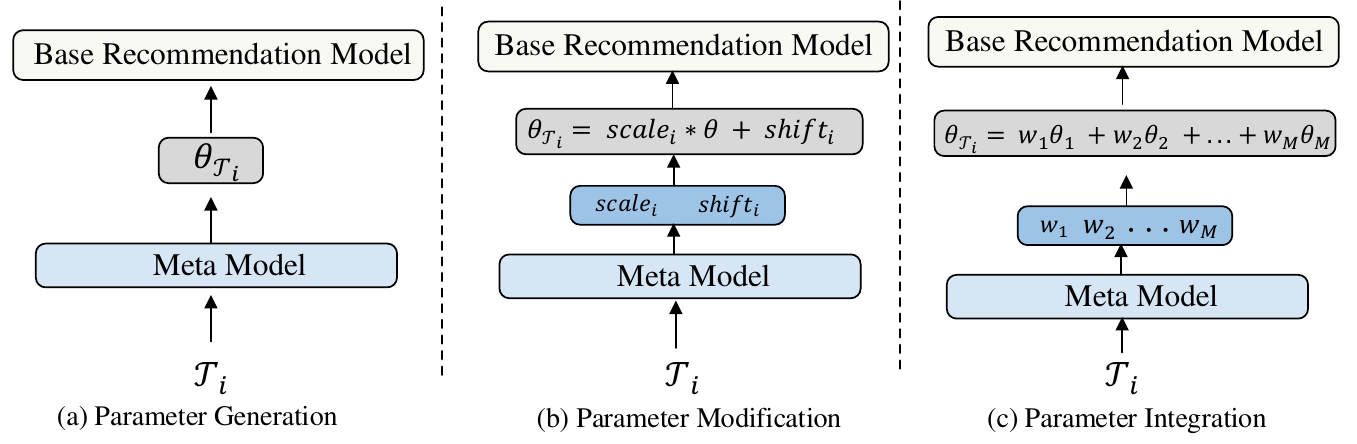}
	\caption{Illustration of different parameter modulation strategies including parameter generation, parameter modification, and parameter integration. One basic example is presented for each category.}
	\label{img6}
\end{figure}

\textbf{Model-based Parameter Modulation.}
Another category of meta-learning based approaches for cold-start recommendation adopts model-based meta-learning for parameter modulation. The core idea to is train a meta model $\mathcal{F}_\omega$ which directly controls or alters the state of base recommendation models without relying on inner-level optimization. More specifically, the form of meta model is usually a learnable neural network that takes interactions in the support set of a task and other useful information (such as losses or gradients) as input to learn task-specific information. The ways of altering states of the base model for a task depend on the design of different methods, i.e., the output form of the meta model. For instance, some works adopt parameter-generation strategies, which directly treat the outputs of the meta model as the task-specific parameters of the base model. Meanwhile, some works take more indirect ways such as gating-based modification of globally shared parameters. We summarize three categories of parameter modulation strategies including parameter generation, parameter modification, and parameter integration, which are illustrated in Fig \ref{img6}. Table \ref{tab6} shows the summary of model-based parameter modulation methods.

One strategy for designing meta-models for parameter modulation is to directly generate task-specific parameters of base models. For instance, Vartak et.at, \cite{LWA} propose two models named \textbf{LWA} and \textbf{NLBA}, to address item cold-start problem. Both LWA and NLBA adopt similar deep neural network architectures as meta models to implement parameter generation strategies. The differences of these two models are the form of recommendation models and parameters to be adjusted. Specifically, take the LWA as the example,  the meta-leaner $\mathcal{F}_\omega$ consists of two sub-networks  $\mathcal{G}(.)$ and $\mathcal{H}(.)$. The first sub-network $\mathcal{G}(.)$ learns task representations based on interacted items of a given user. Embeddings of positive interactions and negative interactions are aggregated as $R^{p}_{i}= \mathcal{G}(I^{p})$ and $R^{n}_{i} = \mathcal{G}(I^{n})$ respectively. The second sub-network $\mathcal{H}(.)$ directly adjusts the base model based on $R^{p}_{i}$ and $R^{n}_{i}$ by learining a vector $\bm{w_i} = \bm{w}_p R^{p}_{i} + \bm{w}_n R^{n}_{i}$. Here, $\bm{w_i}$ are the generated  linear transformation parameters of a logistic regression (LR) function, which is specific for user $u_i$. Then the logistic regression function will act as the user-specific recommendation model to predict the interaction probablity of a new item. Similarly, NLBA utilize a neural network classifier as the base model and generate bias parameters of all hidden layers to implement paramter generation.

To improve tail-item recommendation, i.e., item cold-start recommendation, Zhang et.at \cite{MIRec} propose \textbf{MIREC}, which focuses on transferring knowledge
from head items with rich user feedback to tail items with few interactions. Following the parameter-generation strategy in model-based meta-learning, a meta-mapping module is designed to transfer parameters of a few-shot model to a many-shot model, which achieves the model-level augmentation. Specifically, a meta model $\mathcal{F}_{\omega}$ learns to capture the model parameter mapping from a few-shot model to a many-shot model. The meta-knowledge to be learned in MIREC could be explained as the knowledge about model transformation when more training data are observed. Given a base model $g_\theta$, many-shot model $g_{\theta^*}$ parameterized with $\theta^*$ is learned by feeding all user feedback. Then, to learn meta-knowledge of model transformation, the meta model $\mathcal{F}_\omega$ is incorporated into the training process of a few-shot model $g_{\theta_k}$ (trained with tail items that have less than $k$ interactions) to by minimizing the following objective function: 
\begin{equation} 
\mathcal{L}(\omega, \theta_k) = ||\mathcal{F}_\omega(\theta_k) - \theta^*||^2 + \mathcal{L}_{rec}(g_{\theta_k}, D_k)
\end{equation}
where $\mathcal{F}_\omega(.)$ tasks the parameters $\theta_k$ of the few-shot model as input and generate many-shot model parameters. The first L2 normalization term is utilized to train the parameter mapping ability of $\mathcal{F}_\omega$ from few-shot models and many-shot models. After training, the final recommendation model is obtained by integrating both the original many-shot model $g_{\theta^*}$ and the meta-mapped few-shot model $g_{\mathcal{F}_\omega(\theta_k)}$, in order to perform well on both head and tail items.

Another common strategy of designing meta models for parameter modulation is to modify globally shared parameters into task-specific ones. Instead of directly taking the outputs of meta models as parameters of base models, the core idea of the parameter-modification strategy is to tailor global parameters into task-specific ones under the control of meta models. Lin et.at \cite{TaNP} propose \textbf{TaNP}, which designs a task relevance aware parameter modulation mechanism to customize task-adaptive parameters for base recommendation models. Specifically, TaNP approximates each task as an instantiation of a stochastic process and utilizes an encoder and decoder structure as the preference estimation module, i.e., the base recommendation model. The meta model is designed for modulating parameters of the decoder module. Specifically, the meta model $\mathcal{F}_\omega$ first leverages a  task identity network to encode interactions and a learnable global pool to automatically learn the relevance of different tasks. By doing this, the task representation is obtained as $\bm{o}_i$ and utilized to provide task relevance aware information for parameter modulation. Two candidate modulation strategies including FiLM \cite{FILM} and an extended Gating-FiLM are discussed to scale and shift the parameters of hidden layers of the decoder. Take the FiLM as an example, for the user $u_i$ the adjustment of the $l$-th hidden layer can be defined as: 
\begin{eqnarray}
scale_i^l = tanh(\bm{W}_a^l \bm{o}_i),  shift_i^l = tanh(\bm{W}_b^l \bm{o}_i),
\\
\bm{x}^{l+1}_{i} = ReLU(scale_i^l \odot (\bm{W}_{dec}^{l} \bm{x}^{l}_{i} + \bm{b}_{dec}^{l}) + shift_i^l)
\end{eqnarray}
where $\bm{W}_{dec}^{l}$ and $\bm{b}_{dec}^{l}$ are global parameters of the encoder. $scale_i^l$ and $shift_i^l$ are generated model modulation signals by the meta model. $\bm{x}^{l}_{i}$ is the inputs of the $l$-th layer of the decoder. In this way, TaNP achieves the task-adaptive parameter modulation leveraging model-based meta-learning. 

Similar parameter-modification strategy is also utilized in \textbf{CMML} \cite{CMML}, which utilizes two context encoders and a contextual modulation network as the meta model. Specifically, these two context encoders focus on extracting task context information of the cold-start tasks at the task level and instances(or interaction)-level, respectively. Then, the final context representation $c_{u_i,v_j}$ will be inputted as a hyper-network to generate modulation weights for the specific interaction $r_{u_i, v_j}$. Three network modulation strategies provided by CMML is \emph{Weight Modulation}, \emph{Layer Modulation} and \emph{Soft Modulation}. Specifically, weight modulation only generates weights and bias for the final linear layer. Layer modulation follows FilM and generates weights and bias for linear modulation on layers' output similar to equation (19-20). Soft Modulation is conducted by introducing mixture of experts networks to generate dynamic routing weights for aggregating outputs of multiple subnetworks. Details of three network modulation strategies can be seen in the original paper.

Another work \textbf{Heater} \cite{Heater} leverages the parameter-integration strategy in model-based meta-learning for cold-start recommendation. By incorporating auxiliary information of cold-start users and items, Heater mainly transforms user/item auxiliary representations into collaborative filtering (CF) space and estimates the preference probability. They argue that personalized transformations to different users or items are required. Therefore, they propose to adopt a Mixture-of-Experts \cite{mixture-of-experts}to act as the meta-model for implementing personalized user transformation function $f^U_i$ and item transformation function $f^I_j$. Take the user side as an example, the Mixture-of-Experts consists of $M$ parallel experts with the same structure. Each expert $f^m$ takes the user representation $\bm{u}_i$ as input, and outputs a transformed representation $f^m(\bm{u}_i)$ of the user. The parameter-modification strategy works by adaptively combining outputs of all experts $\{f^1(\bm{u}_i),...,f^M(\bm{u}_i)\}$ with learnable weights. This is equivalent to an adaptive integration of the parameters of multiple experts. As a result, the final transformation function $f^U_i$ is user-specific for each user. 

\textbf{Metric-based Embedding Space Learning.} 
Metric-based meta-learning is also utilized in cold-start recommendation to meta-learn embedding space for embedding similarity comparison.  To alleviate cold-start problem in long-tail item recommendation, Sankar et al.\cite{ProtoCF} proposes \textbf{ProtoCF} which learns a shared metric space for measuring embedding similarities between candidate cold-start items and users. Specifically, inspired by the Prototypical Networks \cite{Prototypical_nets}, 
ProtoCF learns to compose discriminative prototypes for tail items from their few-shot interactions. Based on the support set $\mathcal{S}_i$, the prototype representation for each item $v_i$ is first computed as the mean vector of pretrained user embeddings. Then, a fixed number of group embeddings are learned as external memories to enrich prototype representations of each item. Finally, following the framework of metric learning, given a query user, the similarities between prototype representations  $\{\bm{p}_1,...,\bm{p}_N\}$ of candidate items and the user representation $\bm{u}_i$ are computed in the meta-learned metric space. 

Borrowing the idea of measuring embedding similarity,  Hao et al. \cite{PRE-TRAINING} study how to pretrain GNNs to learn embeddings for cold-start users and items via few-shot reconstruction tasks.  Instead of learning embedding space for calculating embedding similarity between users and items, the \textbf{PreTraining} approach focuses on learning reconstruction space for comparing reconstructed embeddings of few-shot users/items and their ground truth embeddings learned from abundant interactions.  Reconstruction tasks first select target users/items that have sufficient interactions and simulate cold-start situations by sampling a few neighbors for each target user/item. Assuming embeddings trained with abundant interactions are ground truths, the goal of the reconstruction tasks is to reconstruct embeddings based on few-shot neighbors. By measuring and maximizing the similarities among the reconstructed embeddings and the ground truths, the pretrained GNNs are supposed to learn effective embedding space for cold-start users and items.

\subsection{Meta-learning in CTR Prediction}
We summarize details of meta-learning methods in click-through rate prediction from three perspectives, i.e., meta-learning techniques, used auxiliary information, and meta-knowledge representations, as shown in Table \ref{tab7}. Next, we will elaborate on two groups of methods including \emph{Optimization-based Item Embedding Initialization} and \emph{Model-based Item Embedding Genetation}. 

\begin{table}[t]
	\caption{Details of recommendation models with meta-learning methods in click through rate prediction.}\label{tab7}
	\centering
	\begin{tabular}{cccc}
		\toprule
		\textbf{Method} & \textbf{\makecell[c]{Meta-learning \\ Technique}} & \textbf{\makecell[c]{Auxiliary \\ Information}} & \textbf{\makecell[c]{Meta-knowledge \\ representation}} \\
		\midrule
		Meta-Embedding \cite{Meta-Embedding} & Optimization-based & Item Attributes & \makecell[c]{Embedding Initialization \\ \& Meta Model} \\
		\cline{1-4}
		DisNet \cite{DisNet} & Optimization-based & Revelant Items &  \makecell[c]{Embedding Initialization  \\ \& Meta Model }\\
		\cline{1-4}
		GME \cite{GME} & Optimization-based & \makecell[c]{Item Attributes \& \\ Relevant Items}  & \makecell[c]{Embedding Initialization  \\ \& Meta Model }\\
		\cline{1-4}
		TDAML \cite{TDAML} & Optimization-based &  Item Attributes  &  \makecell[c]{Embedding Initialization \& \\  Meta Model \& Sample Weights} \\
		\cline{1-4}
		MWUF \cite{MWUF} & Model-based & \makecell[c]{Item Attributes \& \\ Interacted Users} &   Meta Model \\
		\cline{1-4}
		Meta-SSIN \cite{Meta-SSIN} & Optimization-based & Historical Items &  Parameters Initialization \\
		\bottomrule
	\end{tabular} 
\end{table}

\begin{figure}[tb]
	\centering
	\includegraphics[width = 0.95\textwidth]{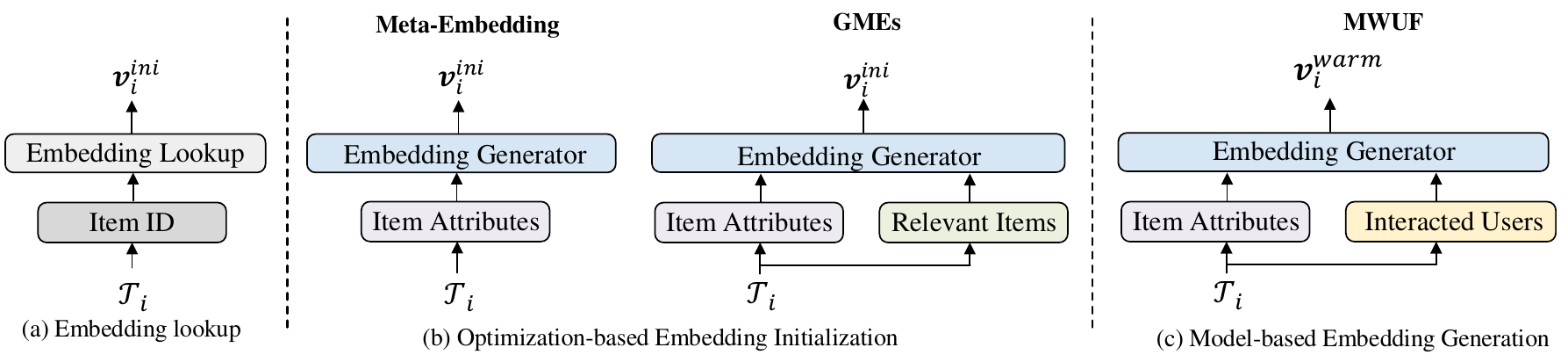}
	\caption{Illustration of different structures of embedding generators in meta-learning methods for CTR prediction. We mainly compare them from what kind of auxiliary information is considered when generating initial embeddings or warm embeddings for new items.}
	\label{img7}
\end{figure}

\textbf{Optimization-based Item Embedding Initialization. } This category of methods mainly focuses on learning initial embeddings for new items, so as to achieve better cold-start and warm-up performance. The main idea of this category is to design an external ID embedding generator as a meta-learner, and apply it to generate adaptive initial ID embeddings for different items newly arrived.  The meta-learner is trained under the optimization-based meta-learning framework.  

Pan et al., \cite{Meta-Embedding} firstly propose the idea of meta-learning a initial embedding generator to replace the randomly intialization strategy for click-through rate prediction problem. Specificially, as shown in Fig \ref{img7} (b), an item/Ad features based embedding generator named \textbf{Meta-Embedding} is designed to take Ad attributes as inputs and generate item-specific initial embeddings $\bm{v}^{ini}_i$. Then the generated user ID embedding $\bm{v}^{ini}_i$ is combined with other feature embeddings such as user embeddings, item attribute embedddings and context embeddings and fed into pretrained predcition models, e.g., DeepFM \cite{Deepfm}, PNN \cite{PNN}, Wide\&Deep \cite{WideDeep}. For the meta-optimization of the Meta-Embedding generator, two batches of labeled instances are sampled for each cold-start item. The first batch $\mathcal{D}_{i}^{a}$ is utilized to evaluate the cold-start performance by directly making predictions with $\bm{v}^{ini}_i$. The second batch $\mathcal{D}_{i}^{b}$ is utilized to evaluate the warm-up performance by making predictions with item embedding $\bm{v}^{warm}_i$ which is locally updated over the first batch data $\mathcal{D}_{i}^{a}$. By doing this, two losses $\mathcal{L}_{cold}(\bm{v}^{ini}_i,\mathcal{D}_{i}^{a})$ and $\mathcal{L}_{warm}(\bm{v}^{warm}_i,\mathcal{D}_{i}^{b})$ are obtained in cold-start phase and warm-up phase, respectively. Based on a unified loss, i.e., $\mathcal{L}_{meta} = \delta \mathcal{L}_{cold} + (1-\delta)\mathcal{L}_{warm}$, outer-level update of optmization-based meta-learning is performed to globally optimize the generator through gradient descent.

Following the idea of item ID embedding generation, several works mainly extend the forms of embedding generators by leveraging other auxiliary information besides item attributes, especially information from relevant users and relevant items. Ouyang et al. \cite{GME} propose a series of graph meta embedding (\textbf{GMEs}) models to learn initial item embeddings based on not only item attributes but also existing relevant items. As shown in Fig \ref{img7} (b), GMEs first connect existing items with new items with graphs through shared item attributes and then apply the graph attention networks to distill neighborhood information for generating embeddings of cold-start items. Three different strategies for distilling information from existing items including pre-defining item embeddings, generating item embeddings from item attributes, and directly aggregating attribute embeddings without learning ID embeddings, are discussed in different variants of GMEs.  Similar to the Meta-Embedding, GMEs also resort to the optimization-based meta-learning framework to train the graph neural network based embedding generator with two sampled batches of each task. Similarly, Li et al. \cite{DisNet} proposes a deep interest-shifting network \textbf{DisNet} which includes a meta-Id-embedding generator (RM-IdEG) module as the initial ID embedding generator. RM-IdEG mainly collects a set of existing items relevant to the target cold-start item through item relations and learns an attentional representation as to the initial ID embedding. Similar to the Meta-Embedding, the optimization of RM-IdEG is separated from pretraining the whole DisNet model and conducted by minimizing both cold-start loss and warm-up loss with optimization-based meta-learning.

Under the framework of optimization-based item embedding initialization, i.e. Meta-Embedding, the optimization strategy is also studied to improve the adaptation ability against the diversity of the task difficulty. Cao et al. \cite{TDAML} proposed a task-distribution-aware meta-learning method (shorted as \textbf{TDAML}) to ensure the consistency between the loss weight and task difficulty when globally updating the embedding generator. They argue that different tasks should have different difficulties in the meta-training phase and assigning equal weights to all tasks may pay limited attention to the hard tasks. On top of the meta-embedding framework, TDAML proposes to adaptively assign different weights when summing the meta losses of different tasks. By modeling the weights $\bm{p}_i$ of meta-losses as the description of task difficulty, extra constraints expecting strong consistency between  $\bm{p}_i$ and meta-loss of the task, i.e., $\mathcal{L}_{meta}^{i}$, are added to find an adaptive loss weight which replaces the uniform weight. As a result, the meta-optimization phase could pay more attention to the harder tasks and achieves better performance improvement.

\textbf{Model-based Item Embedding Generation. } Besides optimization-based techniques, model-based meta-learning is also applied to generate initial item embeddings for better click-through rate prediction performance. Zhu et al. \cite{MWUF} propose \textbf{MWUF} which aims to meta-learn scaling and shifting functions for generating ID embeddings of cold-start items. As shown in Fig \ref{img7} (c), different from optimization-based item embedding initialization above, MWUF directly transforms the cold item ID embedding $\bm{v}^{cold}_i$ of the item $v_i$ to a warm item ID embedding $\bm{v}^{warm}_i$ by applying a scaling and shifting function as follows:  
\begin{eqnarray}
\bm{v}^{warm}_i = \bm{v}^{cold}_i \cdot h^{scale}(\bm{x}_i) + h^{shift}(\bm{U}_i),
\end{eqnarray}
where $\bm{x}_i$ denotes the item feature embedding of the item $v_i$ and $\bm{U}_i$ denotes embeddings of its interacted users. Here a meta scaling network $h^{scale}(*)$ takes $\bm{x}_i$ as input and generate personalized scaling parameters. A meta shifting network $h^{shift}(*)$ takes $\bm{U}_i$ as input and generate personalized shifting parameters. After obtaining the warm ID embedding $\bm{v}^{warm}_i$, MWUF directly make predictions based on pretrained recommendation models such as Wide\&Deep \cite{WideDeep}, DIN \cite{DIN} and AFM  \cite{AFM}. The meta models, i.e., two meta networks, are optimized by minimizing the warm loss, which is obtained by making predictions with  $\bm{v}^{warm}_i$ over observed interactions of the item $v_i$.

\subsection{Meta-learning in Online Recommendation}
In practical large-scale recommender systems, new interaction data are collected continuously. Therefore, newly arrived data should be leveraged to update the recommendation models timely, so as to capture evolving preference trends. Meta-learning methods are also studied in such online settings in order to enhance the ability to efficiently update recommendation models.  Table \ref{tab8} summarize meta-learning based methods in online recommendation from three perspectives. Next, we will elaborate on three groups of methods which are divided according to different levels of modeling updating in the following.

\begin{table}[t]
	\caption{Details of recommendation models with meta-learning methods in online recommendation.}\label{tab8}
	\centering
	\setlength{\tabcolsep}{3mm}
	\begin{tabular}{cccc}
		\toprule
		\textbf{Method} & \textbf{\makecell[c]{Meta-learning \\ Technique}} & \textbf{\makecell[c]{Task Division}} & \textbf{\makecell[c]{Meta-knowledge \\ representation}} \\
		\midrule
		S2Meta \cite{S2Meta} & Optimization-based & Scenario-specific & \makecell[c]{Parameters Initialization \& \\  Meta-learner \& Hyperparameter} \\
		\cline{4-4}
		FLIP \cite{FLIP} & Optimization-based & Sequence-specific &  Parameters Initialization \\
		\cline{4-4}
		FORM \cite{FORM} & Optimization-based & User-specific & \makecell[c]{Parameter Initialization \\ \& Hyperparameter}\\ 
		\cline{4-4}
		SML \cite{SML} & Model-based &  Time-specific &  \makecell[c]{Meta Model} \\
		\cline{4-4}
		ASMG \cite{ASMG} & Model-based & Time-specific &  Meta Model \\
		\cline{4-4}
		LSTTM \cite{LSTTM} & Optimization-based & Time-specific & \makecell[c]{Parameter Initialization}\\
		\cline{4-4}
		MeLON \cite{MeLON} & Model-based & Time-specific & \makecell[c]{Meta Model  \& Hyperparameter}\\
		\bottomrule
	\end{tabular} 
\end{table}

\textbf{User-level Preference Updating.} 
This group of methods mainly divides new interactions according to different users, and designs online learning strategies to learn dynamic user preferences changing over time.
Liu et al. \cite{FLIP} propose \textbf{FLIP} which aims to decouple the learning of user intent (i.e., dynamic short-term interest) and preference (i.e., stable long-term interest) by treating user intents of different user sessions as meta-learning tasks. Instead of jointly learning user intent and preference from newly arrived user visit sequences, FLIP separately learns intent embeddings only based on the interactions of the current session while learning the preference embedding of the user during the whole online learning procedure. Specifically, inspired by an optimization-based meta-learning framework under online setting Online MAML \cite{OnlineMAML},  FLIP learns the initial intent embedding for all sessions which is expected to quickly adapt to each new session. The support set of a task consists of the first $m$ interactions in the session, and the rest is treated as the query set. The outer-level update of the initial intent embedding is performed across a batch of tasks. Therefore, by learning user intent embedding with optimization-based meta-learning techniques, FLIP enhances the ability of user-level preference updating, especially capturing short-term preference evolution during the online learning procedure.

Another work \textbf{FORM} \cite{FORM} also studies meta-learning-based online recommendation based on user-specific task division. To adapt the optimization-based meta-learning to fluctuating online scenarios, FORM enhances the MAML framework to provide a more stable training process in the following directions. First, during local updates of current interactions of a user, a follow the online meta leader (FTOML) algorithm is designed to preserve prior knowledge extracted from all historical interactions of the user. In this way, the updated model during the online training procedure is expected to perform well on not only current data but also prior data, which stables user preference learning. Second, to ensure a consistent update process, a regularized term is added to the loss function to restrict the model parameters as sparse. Third, considering that users with abundant interactions have fewer fluctuations, FORM is designed to assign larger learning rates to users who have larger record lengths and smaller variance of gradients. With the three designs for tackling the fluctuating and noisy nature of online scenarios, FORM is expected to provide a more stable meta-optimization phase for online recommenders.

\textbf{Scenairo-level Model Updating.} Besides conducting user-level preference learning, Du et al. \cite{S2Meta} considers scenario-specific recommendation tasks and proposes a sequential meta-learner \textbf{S2Meta} to automatically learn personalized models for newly appeared scenarios. For instance, scenario-specific tasks could be defined according to item category, item tag, theme events, and so on.  When a small size of interactions are collected online in a new scenario $s_i$, S2Meta aims to fastly update an initial base model $f_\theta$ to a scenario-specific recommendation model $f_{\theta_i}$. Specifically, the meta-knowledge to be globally learned is defined as three factors controlling the inner-level learning, including initial parameters, learning rates, and early-stop policy. The local update of each recommendation task is considered as a sequential learning process consisting of initializing, finetuning with adaptive learning rates, and stopping timely. The sequential learning process is automatically controlled under three parts of a designed meta model which is learned under the optimization-based meta-learning framework.

\textbf{System-level Model Retraining.} Online recommendation systems usually require periodical model retraining with new instances to capture current trends effectively.  Recently, several works formalize the model retraining tasks from the perspective of meta-learning and study meta-learning based model retraining in the online recommendation \cite{SML, ASMG, LSTTM, MeLON}. 

Zhang et.al \cite{SML} firstly investigate the model retraining mechanism from the scheme of meta-learning. At a time period $t$, the model retraining task $\mathcal{T}_{t}$ is constructed with interactions $D_t$ collected currently as the support set, and interactions $D_{t+1}$ in the next time period as the query set. The goal of the model retraining task $\mathcal{T}_{t}$ is to incrementally update the recommendation model $f_{\theta_{t-1}}$ obtained in the $t-1$ time period to a new one $f_{\theta_t}$ which is expected to achieve better performance in the next time period, i.e., $t+1$.  Zhang et al. apply model-based meta-learning techniques to directly transfer parameters $\theta_{t-1}$ to model parameters $\theta_{t}$ with a meta model. Specifically, the meta model utilizes convolutional neural networks as a transfer component which inputs previous parameters $\theta_{t-1}$ and  parameters $\hat{\theta_{t}}$ that are locally updated over $D_t$. The parameters of the next recommendation model $f_{\theta_{t}}$ are generated from the outputs of the transfer component. To make the learned model serve well in the next time period, the loss over $D_{t+1}$ is observed to update the parameters of the meta model. Since the meta-learning based model retraining framework above is operated in a sequential manner, thus the method is named Sequential Meta-Learning (\textbf{SML}).

Following the idea of SML, Peng et al. \cite{ASMG} propose another model retraining method \textbf{ASMG}, which is devise to generate the current model $f_{\theta_{t}}$ based on a sequence of historical models $\{f_{\theta_{1}},...,f_{\theta_{t-1}}\}$. Different from SML, ASMG replaces the CNN-based transfer module with gated recurrent units (GRU) as a meta-generator that captures long-term sequential patterns in model evolution. The meta generator inputs a truncated sequence of historical models of previous periods sequentially. Then the final hidden state $\bm{h}_t$ of the GRU is transformed to generate the parameters of current model $f_{\theta_{t}}$. Similar to SML, the meta-generator in ASMG is also optimized towards better performance over interactions of the next time period $t+1$.

Different from SML which focuses on updating parameters based on the whole data in the current time, one up-to-date approach \textbf{MeLON} \cite{MeLON} further distinguishes the importance of different interactions in the data of the same time. Specifically, given an interaction $r$,  MeLON aims to learn a adaptive learning rate $\alpha_{r,m}$ for $m$-th dimension $\theta_{t}^m$ of current model parameters $\theta_{t}$. A meta model is designed to generate the adaptive learning rate based on information from both the interaction (e.g., relevant historical interactions) and the parameter (e.g. loss and gradient). By assigning adaptive learning rates for each interaction-parameter pair, MeLON hopes to be able to update recommendation models more flexibly in online scenarios.

Besides the model-based meta-learning techniques above, model retraining is also studied under the optimization-based meta-learning framework. Xie et al. \cite{LSTTM} propose \textbf{LSSTM} for online recommendation, which relies on graph neural networks based recommendation models to extract user short-term and long-term preferences. Considering the dynamic nature of short-term preferences in online scenarios, LSTTM constructs model retraining tasks according to different time periods and applies optimization-based meta-learning to learn better initialization of a short-term graph module. Instead of training only based on current data with meta-learning, the global long-term graph module is trained constantly during the whole online learning phase. In this way, short-term preference for new trends or hot topics is captured timely from the recent interactions while long-term preference which reflects users' stable interests is also maintained after the model retraining.

\subsection{Meta-learning in Point of Interest Recommendation}  As shown in Table \ref{tab9}, we summarize meta-learning based methods in POI recommendation from three perspectives, i.e., task division and sequential information, and meta-knowledge representations.  Next, we will elaborate on two groups of methods that study optimization-based sample reweighting and optimization-based parameter initialization, respectively.

\begin{table}[t]
	\caption{Details of recommendation models with meta-learning methods in POI recommendation.}\label{tab9}
	\centering
	\setlength{\tabcolsep}{3mm}
	\begin{tabular}{cccc}
		\toprule
		\textbf{Method} & \textbf{\makecell[c]{Task Division }} & \textbf{\makecell[c]{Sequential \\ Information}} & \textbf{\makecell[c]{Meta-knowledge \\ representation}} \\
		\midrule
		PREMERE \cite{PREMERE} & User-specific & Sequential-free & \makecell[c]{Meta model \& \\ Sample Weight} \\
		\cline{4-4}
		MFNP \cite{MFNP} & User-specific & Sequential-aware &   \makecell[c]{Parameter Initialization} \\
		\cline{4-4}
		CHAML \cite{CHAML} & City-specific & Sequential-aware & \makecell[c]{Parameter Initialization \\ \& Sample Weight } \\
		\cline{4-4}
		Meta-SKR \cite{Meta-SKR} & User-specific &  Sequential-aware & 
		\makecell[c]{Parameter Initialization \\ \& Meta Model} \\
		\cline{4-4}
		MetaODE \cite{MetaODE} & City-specific & Sequential-aware &  \makecell[c]{Parameter Initialization} \\ 
		\bottomrule
	\end{tabular} 
\end{table}

\textbf{Optimization-based Sample Reweighting.} Due to the sparse and noisy nature of check-in data, it is beneficial to assign higher weights to effective instances for better model training. Considering that harder tasks have higher values for boosting model performance, Chen et al. \cite{CHAML} proposes a meta-learning framework \textbf{CHAML} for net POI recommendation which incorporates hardness-aware sampling into optimization-based meta-learning. This work focuses on extracting meta-knowledge from existing cities with sufficient data to cold-start cities with limited check-in instances. By treating POI recommendation in each city as a task, CHAML extends the MAML framework to learn the initial weights of an attention-based sequential recommendation model in order to quickly adapt to cold-start cities. For enhancing the efficiency of model training, the idea of hardness-aware sampling is to sample difficult tasks which have low accuracies. Specifically, the batch of training tasks are not sampled randomly but conditioned on the difficulties of different users and different cities. When generating each task batch, both city-level hardness and user-level hardness are considered via two sampling steps. For the first step, with a group of hard tasks $\mathcal{T}_{hard\_city}$, some hardest users with the lowest prediction accuracies are kept and others are re-sampled to form a new batch of tasks $\mathcal{T}_{hard\_user}$ with harder users. For the second step, a step of the global update is performed over $\mathcal{T}_{hard\_user}$ and then another batch of tasks  $\mathcal{T}_{hard\_city}$ are constructed by keeping some harder cities with lower accuracies and resampling others. In addition, curriculum learning is adopted to measure city-level difficulties with a pretrained teacher so as to generate an easy-to-hard training curriculum. 

Another work \textbf{PREMERE} proposes an adaptive reweighting scheme based on model-based meta-learning in the POI recommendation problem. A meta model $\mathcal{F}_\omega$ is designed to generate sample weights which induce the learning phase of the recommendation model to focus more on valuable samples. The generated weight $w_i=\mathcal{F}_\omega(\bm{x}_i)$ is utilized as the weight of loss summation during recommendation model training. Specifically, $\bm{x}_i$ represents the context of a sample (e.g., user visit entropy, geographical similarity, and temporal similarity) and its loss obtained by the recommendation model. In this way, samples justified as more effective for model training could be adaptively assigned higher weights.  Different from CHAML which evaluates the importance of samples during the sampling phase,  PREMERE randomly sample instances but focuses on reweighting losses of instances in the sampled batch. 

\textbf{Optimization-based Parameter Initialization.} Recently, optimization-based meta-learning methods are also leveraged to learn parameter initialization of specific modules in the next POI recommendation models. Sun et al. \cite{MFNP} propose \textbf{MFNP}, which captures user-specific preferences and region-specific preferences with two LSTM-based modeling modules, respectively. By initializing the parameters of the recommendation model, MFNP locally updates models on corresponding support sets for different users and globally optimizes the initialization via the MAML framework. Another work \cite{Meta-SKR} proposes a sequential knowledge graph based recommendation model \textbf{Meta-SKR} for the next POI recommendation. By jointly modeling sequential, geographical, temporal, and social information with designed sequential knowledge graphs, the next POI recommendation problem is considered as a link prediction based on graph embedding learning. To alleviate the check-in sparsity problem in embedding learning,  an optimization-based meta-learning framework LEO \cite{LEO} is adopted to generate the weights of the GRU-based and GAT-based sequential embedding network which learns node embeddings from the sequential knowledge graphs. In addition,  optimization-based meta-learning is also utilized in \textbf{MetaODE} \cite{MetaODE}  to learn parameter initialization across multiple source cities with sufficient data, so as to gain better generalization over data-insufficient cities.


\subsection{Meta-learning in Sequential Recommendation}
Sequential recommendation mainly focuses on modeling user behavior sequences to capture the dynamic evolution of user preferences.  Several recent studies incorporate meta-learning to alleviate the cold-start issues in sequential recommendation scenarios \cite{metaCSR, CBML, MetaTL, Mecos}. 

To tackle the data sparseness issues of new users, HUANG et al. \cite{metaCSR} propose a cold-start sequential recommendation model \textbf{metaCSR} to learn global inItialization of a sequential recommender with the MAML framework. The sequential recommender is designed to have a GCN-based representation learning module for learning user and item representations and a self-attention based sequential modeling module for encoding user interaction sequences. The MAML framework is leveraged to globally learn parameters of the sequential recommender across different sequential recommendation tasks. Each task utilizes the first $K_1$ interactions in the user behavior sequence of a user as the support set and the rest $K_2$ interactions as the query set. For each interaction, the sequential recommender relies on a historical interaction sequence to predict the current item.

Similarly, another work \textbf{CBML} \cite{CBML} also applies optimization-based meta-learning into self-attention based sequential recommendation models. CBML utilizes two self-attention layers to learn sequential transition patterns at both the item level and the feature level. Based on the base sequential recommendation model above, a cluster-based meta-learning framework is designed to transfer meta-knowledge shared across similar sequential/session-based tasks. Specifically, CBML adaptively learns a soft-clustering assignment for each task which is constructed with a session and generates parameter gates to guide cluster-aware initialization of the base sequential recommendation model. Here, CBML simply tailors cluster-aware initialization of a prediction layer and assigns global initialization for the rest modules of the sequential recommendation model including embedding layers and self-attention layers.

Instead of learning sequential patterns with self-attention models, Wang et al.\cite{MetaTL} propose a \textbf{MetaTL} framework on top of a transition-based sequential recommendation architecture.  To capture short-range transition dynamics from sequences with limited interactions of cold-start users, MetaTL resorts to the idea of transition-based recommendation. The sequential recommendation task for a cold-start user is formulated to predict the tail item $i_{t+1}$ in a transition pair $\{i_{t }\rightarrow i_{t+1}\}$ (i.e., the query set) given previous transition pairs $\{i_{j}\rightarrow i_{j+1}\}_{j=1}^{t-1}$ (i.e., the support set). The transition-based recommendation model aggregates the trainstional information of the user $u_i$ based on multiple pairs in the support set to obtain a relation representation $\bm{r}_{u_i}$ and calculates the preference score as $-||\bm{i}_{t}+\bm{r}_{u_i}-\bm{i}_{t+1}||^{2}$. MetaTL also applies MAML framwork to learn effective global intialization of the transition model for all cold-start users.

Different from applying optimization-based meta-learning to learn suitable initialization of sequential model, metric-based meta-learning is also studied in the cold-start sequential recommendation scenario. Zheng et al.\cite{Mecos} propose \textbf{Mecos} to address the item cold-start issue in the sequential recommendation. They firstly construct $N$-way $K$-shot classification task by sampling K sequences for N cold-start items, respectively.  Then, Mecos learns holistic representations for support sets and query sets of different items and 
leverages a matching network to calculate the similarity scores between each support and query pair, so as to generate classification results of the $N$ query sets according to the similarity metric. The matching network is optimized in the meta-training phase with constructed classification tasks and could be directedly utilized to make predictions without local adaptation over meta-testing tasks.

\subsection{Meta-learning in Cross Domain Recommendation}
Cross-domain Recommendation (CDR) which aims to transfer knowledge from an informative source domain to the target domain is a promising solution to alleviate the cold-start problem. Several studies  \cite{TMCDR, PTUPCDR} introduce meta-learning into cross-domain recommendation methods to achieve better knowledge transfer under cross-domain settings by extracting prior knowledge.

Under the framework of Embedding and Mapping methods for CDR (EMCDR \cite{EMCDR}) which explicitly learns representation mapping function based on overlapping users, Zhu et al. \cite{TMCDR} propose a transfer-meta framework \textbf{TMCDR} to enhance the training process of EMCDR-based methods. Specifically, similar to the embedding step in the general EMCDR framework, TMCDR firstly learns domain-specific embedding models for both source and target domains, respectively. The idea of meta-learning is utilized in a meta stage, which trains a meta network to transform source embeddings into the target feature space.  In the meta stage, TMCDR samples two groups of overlapping users to construct meta-training tasks which utilize one group as a support set and another group as a query set. The meta network is optimized across tasks under the framework of optimization-based meta-learning. Compared with the original mapping function of EMCDR, the meta network is supposed to have better generalization when transforming user embeddings from a source domain for cold-start users in the target domain.

Instead of applying an optimization-based framework, another work \textbf{PTUPCDR} \cite{PTUPCDR} proposes to directly generate user-personalized bridge functions with a meta network. Following the similar idea of mapping-based knowledge transfer, PTUPCDR also focuses on transferring user preferences from an informative source domain to a sparse target domain. Different from learning a common mapping function for all users, this work considers that the preference transfer should be personalized. Specifically, for a user $u_i$, the personalized parameters $\bm{w}_{u_i}$ of the mapping function are generated with a meta network, in order to transform a user embedding in the source domain $\bm{u}^s_i$ to an initial user embedding in the target domain $\bm{u}^t_i$. The meta network takes representations of users' personalized characteristics which are extracted from user interactions in the source domain as inputs and generates $\bm{w}_{u_i}$ as the parameters of a mapping function. By applying the personalized mapping function on the embedding transfer for the user $u_i$, $\bm{u}^t_i$ could be utilized for predictions on the target domain. The optimization procedure across different cross-domain recommendation tasks enables the meta network to learn the meta-knowledge about personalized parameter generation.

\subsection{Meta-learning in other Recommendation Scenarios}
Besides the recommendation scenarios mentioned above, We will briefly discuss some typical scenarios else, including Multi-behavior recommendation,  Knowledge graph based recommendation, and Recommendation Model Selection. Other sporadic works involving federated recommendation \cite{MetaMF}, size and fit recommendation \cite{Meta-SF}, audience expansion in recommendation \cite{MetaHeac} and interactive recommendation \cite{NICF} will not be presented in detail.

\subsubsection{Multi-behavior recommendation} Multiple types of user behaviors (e.g., click, add-to-cart and purchase) are considered to be able to reflect multi-view user preferences in real-world scenarios. Multi-behavior recommendation aims to capture multi-typed behavior patterns and comprehensively learn users' preferences from their diverse behaviors.

Although previous studies have made efforts to learn complex dependencies among different types of behaviors, two recent works \textbf{MB-GMN} \cite{MB-GMN} and \textbf{CML} \cite{CML} argue that the multi-behavior patterns should be diverse and personalized for different users. Therefore, both of them study the multi-behavior recommendation problem with the meta-learning paradigm.
Specifically, by applying model-based meta-learning, MB-GMN designs two meta networks to directly generate personalized parameters of different users for both a multi-behavior pattern representation learning module and a prediction module. The former meta-network generates personalized weights of behavior-specific context projection layers by taking 
user-specific behavior characteristics as input. The latter meta-network generates personalized parameters of final prediction networks by encoding the target user-item pair as the state representation of the current instance. Following the similar idea of model-based parameter generation, CML leverages a meta weight network to generate personalized weights for integrating contrastive losses in different behavior views. By generating the weighting function based on user-specific behavior characteristics, the meta weight network is designed to adaptively customize the contrastive learning phase for different users.

\subsubsection{Knowledge Graph based Recommendation} To tackle the cold-start problem in knowledge graph based recommendation, Du et al. \cite{MetaKG} firstly attempt to incorporate an optimization-based meta-learning paradigm to simultaneously derive prior knowledge from both collaborative information in interactions and semantic information in knowledge graphs. Specifically, a graph attention network based recommendation model \textbf{MetaKG}, which aggregates information of neighboring entities in a collaborative knowledge graph to learn user and item representations, is utilized as the base model. Then the parameters of the base model are optimized through an optimization-based meta-learning schema. Specifically, the parameters of the base model are divided into a knowledge-aware part and another collaborative-aware part and optimized in different strategies. For the knowledge-aware part involving entity representation learning,  parameters are globally optimized to learn shared semantic information of the whole knowledge graph. Differently, the collaborative-aware part involving preference aggregation is first locally adapted to each task and then globally optimized across different tasks, so as to ensure fast adaptation over cold-start users by learning effective global initialization.

\subsubsection{Recommendation Model Selection}
In practical recommendation systems, a single model is unlikely to always achieve the best performance over every dataset \cite{SelectingCF} or every user \cite{MetaSelector}. Recommendation model selection is a realistic solution, which aims to suitably select or combine different recommendation models in different scopes by discovering relationships between data characteristics and model performance. In previous works \cite{SelectingCF, Selecting_2, Selecting_3}, meta-learning has been understood as a kind of methodology that extracts diverse forms of meta-features from given datasets and induces meta models to predict the best recommendation model based on these meta-features. This line of methods heavily relies on manual extraction of meta-features and thus is out of the range of deep meta-learning that we discussed in this survey. More related works could be found in \cite{ren2019survey, Survey_selection}.

Recently, Luo et al.\cite{MetaSelector} have studied recommendation model selection problem under the framework of optimization-based meta-learning. Given a collection of recommendation models, a model selector \textbf{MetaSelector} is designed to adaptively ensemble all models by generating soft selection weights. By regarding each task as learning suitable model selection weights for a user,  the model selector is optimized across different model selection tasks under an adaptive learning rate augmented MAML framework. In the local adaptation phase, for each task, the model selector is first locally updated with the support set of the user and then
generate personalized model selection weights to evaluate its effectiveness over the query set. In the global optimization phase, the initialization of the model selector is updated across multiple tasks to make sure fast adaptation to new model selection tasks. Note that these recommendation models should be pretrained with all data and kept fixed in the meta-training phase.

\section{Future Directions}
\label{sec:6}
In this section, we analyze the limitations of existing deep meta-learning based recommendation methods and outline some prospective research directions which worth exploring in the future. 

\subsection{Meta-Overfitting } Generalization across different tasks is the key capacity of meta-learning, and it mainly depends on how well meta-learners fit the whole task distribution with meta-training tasks. Similar to overfitting over training instances in conventional machine learning, the meta-overfitting issue occurs when meta-learners merely memorize all meta-training tasks but fail to adapt to novel tasks (i.e, meta-testing tasks) \cite{yin2019meta}. Since the number of training tasks is usually much smaller than the number of instances, the meta-overfitting problem is more severe in meta-learning compared with regular supervised learning \cite{ML_survey_2}. In the field of recommendation systems, existing meta-learning methods mainly construct a fixed and limited number of tasks as summarized in section 4, and thus are likely to suffer from meta-overfitting over meta-training tasks. One straightforward strategy against meta-overfitting is conducting task augmentation during task construction. For instance, for constructing typical few-shot classification tasks, $N$ classes are randomly sampled and $K$ instances of each class are also randomly sampled. In this way, not only the volume of available tasks is greatly increased, but also these tasks are kept mutually exclusive. Some other efforts of task augmentation \cite{zhu2022towards, liu2020task,murty2021dreca}, meta-regularization \cite{yin2019meta} and Bayesian meta-learning \cite{yoon2018bayesian} are also studied and proven effective in addressing the meta-overfitting issue. Therefore, it is a promising direction for developing meta-learning based recommendation models with better meta-generalization abilities.

\subsection{Task Heterogeneity} 
The majority of meta-learning methods adopted in recommendation models mainly focus on globally learning meta-knowledge across different tasks without considering the task heterogeneity problem. However, globally learned meta-learners usually perform well when task distribution is uni-modal, but lack the ability to provide desirable prior knowledge to heterogeneous tasks from the multi-modal distribution \cite{vuorio2019multimodal}. Considering the huge differences from the perspectives of both user interests and item attributes in recommender systems, the distributions of user-specific tasks or item-specific tasks are often complex. Moreover, different from image or NLP tasks, the distribution of recommendation tasks shows strong dynamics with time evolving. Therefore, properly handling the task heterogeneity is essential for learning high-quality meta-knowledge across different tasks. In recommendation systems, several recent works \cite{MAMO, TaNP, PAML}  have explored the task heterogeneity issue under user cold-start scenarios. They mainly trigger user-specific adjustments to the globally shared knowledge (e.g. initialization or parameters modulation) conditioned on the user profile information or interaction information. On this basis, more efforts on how to effectively distinguish different tasks under diverse task distributions are desired. Recent research resorts to more reasonable task clustering structures, such as hierarchical structure \cite{yao2019hierarchically} and meta-knowledge graph \cite{yao2019automated}, to capture complex relations between tasks. In addition, external domain knowledge (e.g., knowledge graphs on the item side or social networks on the user side) also could be incorporated to facilitate identifying task relationships \cite{suo2020tadanet}. Besides, task heterogeneity in other recommendation scenarios such as online recommendation and POI recommendation is also worth exploring.

\subsection{Task Augmentation with Auxiliary Information}
Recent meta-learning based recommendation models mainly leverage interaction data as the information source to construct meta-learning tasks. In practical, the data in recommendation systems could be diverse and multi-modal. Data from other sources (e.g., knowledge base, social networks, user/item side information, cross-domain information) and different modalities (e.g., video, image, and text) could be incorporated to provide auxiliary information. Besides simply enhancing user/item representations by inputting auxiliary data into the base recommendation model, another possible strategy is to perform task augmentation with auxiliary information in order to enrich the context of tasks. One relevant work \cite{MetaHIN} is proposed to incorporate multifaceted semantic contexts into tasks by extending both support set and query set based on the item attribute information. From the user side, the user social network is also utilized to extract preference information from friends, which implicitly augment user-specific tasks \cite{PAML}. Therefore, we believe that developing new-type task construction beyond interaction data not only injects auxiliary information to alleviate data insufficiency issue but also provides motivation for designing novel meta-learning methods from the level of task construction. Meanwhile, in recommendation scenarios where own rich auxiliary information but have not yet been widely studied under the meta-learning paradigm, e.g., knowledge graph based recommendation, review-based recommendation, and cross-domain recommendation, it is necessary to design appropriate meta-learning tasks according to the characteristics of auxiliary information such as structural information, textual information, and cross-domain information.

\subsection{Neural Network Architecture Search for Reocmmendation Models}
Neural network architecture search (NAS) \cite{NAS_survey} is also a popular application where meta-learning techniques have been well studied in computer vision and natural language processing domains. Recent meta-learning based NAS methods mainly focus on learning meta-knowledge about specifying an architecture of a neural network for each task. For instance, one representative work \cite{liu2018darts} designed a bilevel optimization to solve the network architecture search problem for image classification. As a result, task-specific neural architecture could be adapted to each task from a general meta-architecture.  While meta-learning has been seen as a powerful solution for network architecture search for deep neural networks \cite{lian2019towards, Meta_NAS, Meta_NAS_2, Meta_NAS_3, Meta_NAS_4}, the architecture search of neural recommendation models has not been well studied. The most relevant work \cite{MetaSelector}  is closer to the topic of meta-learning about recommendation model selection. For developing meta-learning based NAS for recommendation models, two key points are search space and search strategy. Considering the neural structure of popular recommendation models, the search space might involve FNN structures, RNN structures, CNN structures, and attention-based structures. As for the search strategy, both initial conditions of architectures \cite{lian2019towards, Meta_NAS_2} and meta-models for learning task-agnostic representations \cite{Meta_NAS} are studied with meta-learning techniques. In addition, the mutual impact of structural connections and model weights is also proven beneficial to each other for better optimization \cite{Meta_NAS_2}. Thus, designing a neural network architecture search framework to automatically specify recommendation models for different tasks or datasets could be another future direction.

\section{Conclusion}
\label{sec:7}
The rapid development of deep meta-learning methods has propelled the progress in the research field of recommender systems in recent years. This paper provides a timely survey after systematically investigating a large number of related papers in this area. We broke it down into a taxonomy of recommendation scenarios, meta-learning techniques, and meta-knowledge representations. For each recommendation scenario, technical details about how to apply meta-learning are introduced for existing methods. Finally, we point out several limitations in current research and highlight some promising future directions to promote research in meta-learning based recommendation methods. We hope our survey can be beneficial for both junior and experienced researchers in the relative areas.


\bibliographystyle{ACM-Reference-Format}
\end{document}